\documentclass[prb,aps,twocolumn,floatfix]{revtex4}
\usepackage{bm}
\usepackage{graphicx}

\newcommand{\be}{\begin{equation}}
\newcommand{\ee}{\end{equation}}
\newcommand{\bea}{\begin{eqnarray}}
\newcommand{\eea}{\end{eqnarray}}

\begin{document}
\def\tit#1#2#3#4#5{{#1} {\bf #2}, #3 (#4)}

\title{Bond order from disorder in the planar pyrochlore magnet}
\author{O. Tchernyshyov}
\email{olegt@jhu.edu}
\affiliation{Department of Physics and Astronomy, The Johns Hopkins University,
Baltimore, Maryland, 21218}
\author{O. A. Starykh}
\email{oleg.starykh@hofstra.edu}
\affiliation{Department of Physics and Astronomy, Hofstra University, 
Hempstead, New York, 11549}
\author{R. Moessner}
\email{moessner@lpt.ens.fr}
\affiliation{Laboratoire de Physique Th\'eorique de l'Ecole Normale
Sup\'erieure, CNRS-UMR8541, Paris, France}
\author{A. G. Abanov}
\email{alexandre.abanov@sunysb.edu}
\affiliation{Department of Physics and Astronomy, Stony Brook University, 
Stony Brook, New York 11794-3800}
\date{\today}

\begin{abstract}

We study magnetic order in the Heisenberg antiferromagnet on the
checkerboard lattice, a two-dimensional version of the pyrochlore
network with strong geometric frustration.  By employing the
semiclassical ($1/S$) expansion we find that quantum fluctuations of
spins induce a long-range order that breaks the four-fold rotational
symmetry of the lattice.  The ordered phase is a valence-bond crystal.
We discuss similarities and differences with the extreme quantum case
$S = 1/2$ and find a useful phenomenology to describe the bond-ordered
phases.

\end{abstract}

\maketitle

\section{Introduction}

Frustrated magnets have attracted the attention of theorists for
several decades, beginning with the study of the Ising antiferromagnet
on the triangular lattice.\cite{Wannier} More recently, new families
of frustrated magnetic compounds have become available for
experimental studies reviving the interest in their
properties.\cite{SR,Zinkin96,Greedan01} By its very nature, a
frustrated system has an extremely large classical degeneracy of the
ground state.  This degeneracy is very effective in suppressing
classical spin order,\cite{Villain79} thus providing a route to
non-N\'eel (quantum) ground states even for higher-dimensional
systems.  The nature of such ground states is far from obvious: the
aforementioned degeneracy allows for a variety of unusual vacua.
Among the possibilities are bond-ordered states, in which pair
averages $\langle {\bf S}_i \!\cdot\! {\bf S}_j \rangle$, rather than
spins $\langle {\bf S}_i\rangle$ themselves, form a periodic
pattern;\cite{Harris91,Read91} and spin liquids that break no lattice
symmetry but are distinguished by the unusual quantum numbers and
statistics of their excitations.\cite{RVB}

One of the most intensively studied frustrated systems is the
Heisenberg antiferromagnet on the pyrochlore lattice.  It has many
experimental realizations that show rather remarkable magnetic
behavior.  For example, the spinel ZnCr$_2$O$_4$ is the first
frustrated magnet in which zero modes---spin waves connecting
degenerate ground states---have been observed by neutron
scattering.\cite{SH02} Not long ago, it has been
shown\cite{Yamashita00,Tchernyshyov02a} that a coupling between spins
and lattice vibrations leads to a spin-Peierls phase transition in
this manifestly three-dimensional spin system.  This effect, also
observed\cite{SH00} in ZnCr$_2$O$_4$, is classical in nature in the
sense that it is not parametrically small in $1/S$.  Therefore it is
expected to dominate the more subtle quantum effects for large values
of spin.

Effects of frustration in quantum pyrochlore antiferromagnets,
particularly in the limit of a small spin $S$, are drawing quite a bit
of interest.  Finding answers in this case may provide clues to the
unusual behavior of underdoped cuprate superconductors, where
frustration of the spin system is achieved through the motion of doped
charges.  Although replacing dynamic frustration with geometric one
somewhat simplifies the problem, it is still far from trivial.  An
exact solution for $S=1/2$ is not available and is not expected in the
immediate future.  Numerical diagonalizations are hampered by the
quick growth of the Hilbert space with the lattice size in three
dimensions.  Several research groups are attacking the problem from
various solvable limits: large $S$, \cite{Henley-unpub} large 
$N$,\cite{Tchern-unpub} and weakly-coupled spin 
clusters.\cite{Koga01,Tsunetsugu01,Tsunetsugu02,Brenig,Starykh02} 
Because it is not even obvious that extrapolations from these limits
will lead to a converging answer, it seems highly desirable to test
these approaches on a similar problem for which numerical answers are
available.

Most recently, a two-dimensional version of the pyrochlore network,
the checkerboard lattice\cite{Liebmann86} (also known as the planar pyrochlore
and the square lattice with crossings), has become a focus of
analytical\cite{liebschupp,elhajal,Moessner01} and
numerical\cite{Palmer01,Fouet01,Sindzingre02,Berg02} studies.  Lower
dimensionality of this system makes it an easier target for numerical
approaches; at the same time, it has the local coordination of the
pyrochlore lattice: magnetic bonds form a network of corner-sharing
tetrahedra with spins at the vertices.  It is therefore hoped that
studies of the Heisenberg antiferromagnet on the checkerboard lattice
can shed light on the behavior of its three-dimensional analog.

The planar pyrochlore lattice {\em does} differ from the pyrochlore
proper in one fundamental aspect: not all bonds of its tetrahedra are
equivalent---because no symmetry of the lattice turns first neighbors
(horizontal and vertical bonds) into second (diagonal).  Even if the
corresponding exchange couplings $J_1$ and $J_2$ are set equal, spin
correlations between first and second neighbors tend to be different,
as evidenced by both analytical and numerical results.  A lack of such
symmetry compels one to look at the general case with $J_1 \neq J_2$.

In this work, we study the checkerboard antiferromagnet in the limit
of large spin $S$, which allows for a systematic perturbation theory
in powers of $1/S$.\cite{Oguchi} We then compare our answers to the
available numerical results for the opposite limit, $S=1/2$, and find
a simple phenomenology that describes both rather well.  Our
understanding of the checkerboard antiferromagnet has been greatly
helped by three recent ideas: (a) Henley's `gauge symmetry' relating
degenerate collinear ground states in a frustrated
magnet;\cite{Henley-unpub} (b) casting of the problem in terms of
bond---rather than spin---variables, which were introduced a decade
ago by Harris {\em et al.};\cite{Harris91} (c) a realization of the
potential significance of the $q=3$ Potts model to bond-ordered states
on the pyrochlore and checkerboard.  

Because this paper is long and technical, the reader may find it
helpful to peruse an informal introduction to the subject written by
one of us.\cite{Tch-jpcm} That article states our reasons to pursue
the large-$S$ route to strongly frustrated quantum magnets, explains
the challenges of that approach and points out ways to overcome them.
It also contrasts the results obtained for different two-dimensional
analogs of the pyrochlore lattice (the checkerboard is one of them).

\begin{figure}
\includegraphics[width=0.9\columnwidth]{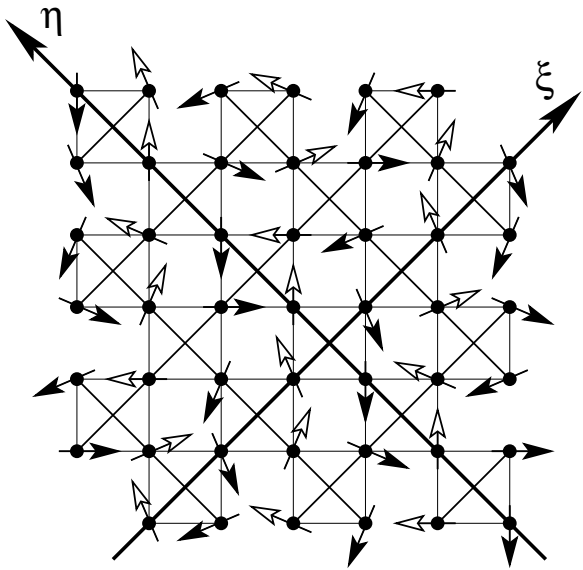} 
\caption{Heisenberg antiferromagnet on the checkerboard lattice.
Shown is a generic classical ground state for a model in which the
second-neighbor coupling $J_2$ that exceeds the nearest-neighbor
coupling $J_1$.  At the level of the classical approximation, the
N\'eel order parameters of individual chains $\hat{\bf n}_i$ are
uncorrelated.  The direction $\hat{\bf n}_i$ is defined as that of
spin ${\bf S}_i$ at the left edge.}
\label{fig-cboard}
\end{figure}

\begin{figure}
\includegraphics[width=\columnwidth]{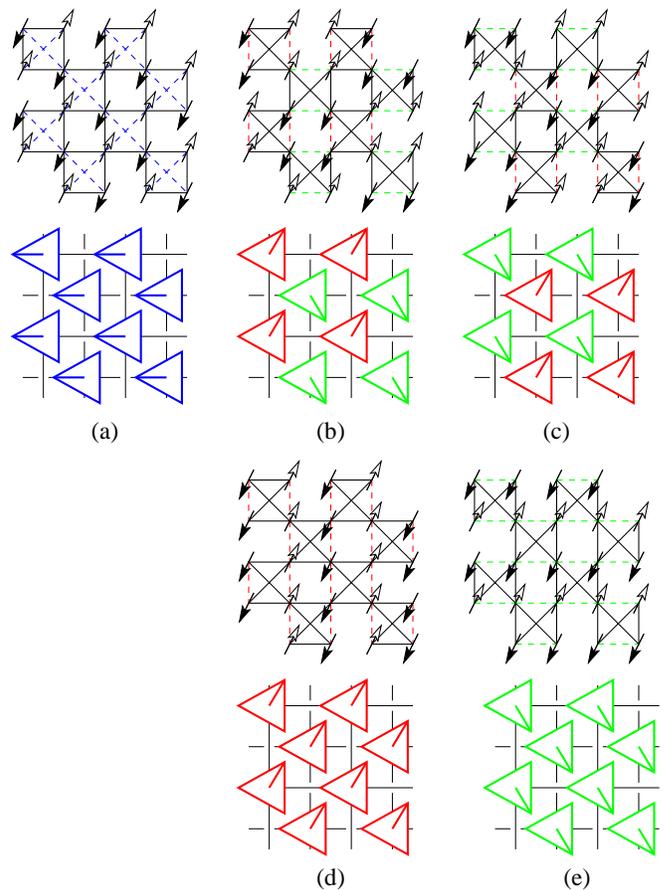} 
\caption{(a) The classical ground state for $J_1>J_2$.  (b) through (e)
are the four classical ground states at the order $1/S$ for $J_1
\leq J_2$.  Frustrated bonds (those with two parallel spins) are shown
in color dashed lines.  Bottom figures show the lattice of tetrahedra;
the primary colors (red, green, and blue) encode the location of
frustrated bonds.  See Fig.~\ref{fig-colors} for more details.}
\label{fig-cboard-vacua}
\end{figure}

The ground state of a classical ($S=\infty$) Heisenberg
magnet is found by minimizing its energy
\begin{equation}
E_0 = \sum_{\langle ij \rangle}J_{ij}\, {\bf S}_i \cdot {\bf S}_j
= {\cal O}(S^2)
\label{E-0}
\end{equation}
with respect to classical spin variables ${\bf S}_i$.  Exchange
coupling is $J_1$ on horizontal and vertical bonds, and $J_2$ along
diagonals.  For weaker diagonal bonds, $J_2 < J_1$, classical energy
minimization gives a unique ground state (modulo a global rotation of
all spins).  The ground state, shown in
Fig.~\ref{fig-cboard-vacua}(a), is the same as that of the simple
square lattice ($J_2 = 0$).

In the region $J_2 \geq J_1$, the classical ground state becomes
continuously degenerate.  For stronger diagonal bonds, $J_2 > J_1$,
the system can be viewed as a collection of criss-crossing chains
running along diagonals of the lattice (Fig.~\ref{fig-cboard}).  Let
us choose north-east to be our positive $\hat{\xi}$-direction, and
north-west to be $\hat{\eta}$.  Classically, each chain has perfect
N\'eel order at zero temperature, however, directions of staggered
magnetizations $\hat{\bf n}_i$ of different chains are completely
independent at the classical level.  In an $L\times L$ lattice with
periodic boundary conditions classical ground states can be
parametrized by $L$ unit vectors $\hat{\bf n}_i$.  The classical
degeneracy increases even further in the case of equal exchanges $J_2
= J_1$.\cite{MCprl,MCprb}

The first-order (in $1/S$) correction to the classical energy comes from the
zero-point quantum fluctuations of spin waves,
\begin{equation}
E_1 = \sum_{k} \hbar |\omega_k|/2 = {\cal O}(S),
\label{E-1}
\end{equation}
where $\{\omega_k\}$ are eigenfrequencies of classical spin waves
about a given ground state obtained from the equations of motion
\begin{equation}
\hbar \dot{\bf S}_i = \sum_{j} J_{ij}\, {\bf S}_i \times {\bf S}_j.
\end{equation}  
It has been established previously that, quite generally,
quantum fluctuations select ground states with collinear spins
(assuming such classical ground states exist).\cite{Shender,Henley87}  

For a fixed global direction $\hat{\bf n}$, the problem thus reduces
to a minimization of the zero-point energy (\ref{E-1}) over a discrete
set of collinear N\'eel states.  Thus selected ground states can be
characterized in the thermodynamic limit with the aid of some order
parameters.  In addition to violating the spin-rotation symmetry O(3),
these ground states can also break some discrete lattice symmetries.
For instance, when the ground states are not symmetric under $\pi/2$
rotations of the plane, one expects an order parameter with the
structure ${\bm Z}_2 \times {\bm S}^2$.  For a given direction of the
N\'eel vector, there should then be {\em two} degenerate ground
states.  (See, e.g., the work by Chandra {\em et al.}\cite{Chandra} on
the square lattice with a large second-neighbor coupling.)  Contrary
to these expectations, we find that for $J_2>J_1$ the ground state is
{\em fourfold} degenerate with an order parameter ${\bm Z}_2 \times
{\bm Z}_2 \times {\bm S}^2$.  The extra degeneracy turns out to be
related to a gauge-like symmetry (due to Henley\cite{Henley-unpub})
that exists at the order $1/S$ in the semiclassical expansion.  This
dynamical symmetry is responsible for an even higher degeneracy of the
ground state at the most frustrated point $J_1=J_2$.  In that case,
the N\'eel order is destroyed and the order parameter is reduced to
${\bm Z}_2 \times {\bm Z}_2 \times {\bm S}^2 / {\bm Z}_2$ (two Ising
orders and a director).  At $J_1=J_2$ the planar pyrochlore is a
valence-bond solid with two independent bond orders and a nematic spin
order.

The paper is organized in the following way.  In most of it (Sections
\ref{sec-1st-order} through \ref{sec-summary}) we study the
lowest-order---${\cal O}(1/S)$---quantum corrections to the degenerate
classical limit.  We first explore the case where the second-neighbor
coupling $J_2$ dominates and the system can be viewed as a set of
weakly coupled antiferromagnetic chains.  The interchain coupling
$J_1$ is frustrated and has no effect at the classical level.  We show
in Sec.~\ref{sec-1st-order} that, in line with the standard
arguments,\cite{Shender,Henley87} quantum fluctuations favor collinear spin
states.  By using $J_1/J_2$ as a small parameter, we derive an
effective interaction between the chains generated by quantum
fluctuations.  This potential is minimized by {\em four} distinct
classical states.  In Sec.~\ref{sec-chains} we prove that this
degeneracy remains intact for all $J_1 < J_2$ and trace its origins to
Henley's gauge symmetry.  Sec.~\ref{sec-cboard} presents our findings
in the strongly frustrated case of equal exchange couplings.  This
time, a much larger degeneracy of the ground state kills the N\'eel
order (replacing it with a nematic order), but the bond order
survives.  A summary of the large-$S$ results is given in
Sec.~\ref{sec-summary}.  Finally, in Sec.~\ref{sec-quantum} we explore
the connection of these large-$S$ results to the $S=1/2$ phase diagram
obtained in numerical studies and speculate on a phenomenology of the
bond order in the $S=1/2$ case.

\begin{figure}
\includegraphics[width=\columnwidth]{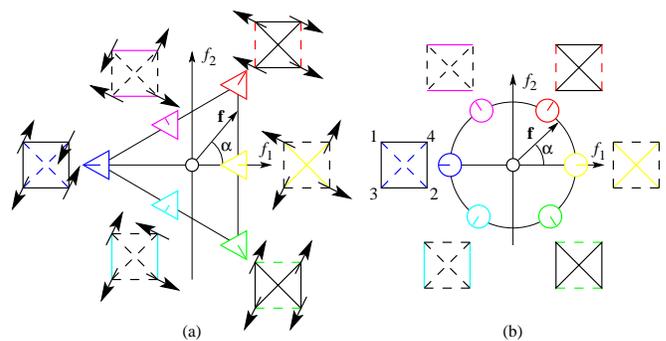} 
\caption{Bond variables ${\bf f} = (f_1, f_2)$ for a single
tetrahedron in a ground state $\sum_{i=1}^4 {\bf S}_i = 0$.  See
Eq.~(\ref{eq:f-def}) for a definition of ${\bf f}$.  (a) Classical
spins.  (b) $S=1/2$.  In both cases, primary colors denote the pure
states with two frustrated bonds (parallel classical spins or a spin
triplet).  Secondary colors mark the pure states with two satisfied
bonds (antiparallel spins or a spin singlet).  The triangle and the
circle delineate the domains of attainable values.  }
\label{fig-colors}
\end{figure}

\section{Weakly coupled chains: $J_1 \ll J_2$}
\label{sec-1st-order}

For $J_1=0$, the magnet is reduced to a collection of independent
antiferromagnetic chains running along the diagonals
(Fig.~\ref{fig-cboard}).  In a classical ground state, each chain has
perfect antiferromagnetic order.  For this reason, there is no
coupling between intersecting chains at the classical level---even in
the presence of a finite interchain coupling $J_1 < J_2$.

Quantum fluctuations disrupt the perfect N\'eel alignment of adjacent
spins and thus enable the chains to interact.  For a weak interchain
coupling $J_1 \ll J_2$, one can use a systematic perturbation theory
in $J_1/J_2$ (staying at the same order in $1/S$) developed by
Shender.\cite{Yildirim}

\subsection{Effective interactions between chains}

To the lowest nontrivial order in $J_1/J_2$, the interchain coupling
generates a potential that selects collinear spin configurations.  
For N\'eel magnetizations of individual chains $\hat{\bf n}_{m}$, 
\begin{equation}
E^{(2)} = 
- \, \frac{4-\pi}{2\pi}\, \frac{J_1^2 S}{J_2} 
\sum_{\text{crossings}}
(\hat{\bf n}_{m} \!\cdot\! \hat{\bf n}_{n})^2.
\label{2order}
\end{equation}
This interaction---coupling any two chains intersecting over a
tetrahedron---is minimized when all staggered magnetizations point
along a common axis, $\hat{\bf n}_m = \pm \hat{\bf n}$.  The tendency
of spins to align is the standard outcome of the order-from-disorder
scenario.\cite{Shender,Henley87} See Appendix for a derivation of
Eq.~(\ref{2order}).

Collinear N\'eel states can be characterized by a collection of Ising
variables: $s_{m} = \pm 1$ for chains running along the direction
$\hat{\xi}$ and $t_m = \pm 1$ for chains along $\hat{\eta}$, see
Fig.~\ref{fig-cboard}.  Given a global orientation of spins
$\hat{\bf n}$, these numbers determine the staggered magnetizations of
individual chains, $\hat{\bf n}_m = s_m \hat{\bf n}$ or $t_m \hat{\bf
n}$, respectively.  To the fourth order in $J_1$ (but still to the
first order in $1/S$) we obtain the correction to the classical energy
of the ground state
\begin{eqnarray}
E^{(4)} &=&
\frac{2J_1^4 S}{ J_2^3} 
\sum_{k=1}^{\infty} \sum_{l=1}^{\infty} \sum_{m} \sum_{n} 
(A_{2k} s_{m}s_{m+2k} 
\label{4order}\\
&+& A_{2l} t_{m}t_{m+2l}
-B_{2k-1,2l-1}\, s_{m}s_{m+2k-1}\, t_{n}t_{n+2l-1} ),
\nonumber
\end{eqnarray}
where the positive coefficients $A_l$ and $B_{kl}$ are computed in the
Appendix.  Note that there is no pairwise interaction between adjacent
parallel chains (nor, for that matter, between any parallel chains an
odd distance apart).  As a result of that, the effective interaction
(\ref{4order}) is invariant under the transformation
\begin{equation}
s_m \mapsto (-1)^m s_m, 
\hskip 5mm
t_n \mapsto (-1)^n t_n,
\label{staggering}
\end{equation}
which flips the spins on every other diagonal chain.

\subsection{Ground states}

The largest term in Eq.~(\ref{4order}) is the interaction of 4 
chains intersecting around an empty plaquette:
\begin{equation}
-\frac{2J_1^4 S}{ J_2^3}
\sum_{m}\sum_{n} B_{1,1}\, s_m s_{m+1}\, t_{n} t_{n+1}
\end{equation}
(see Appendix for details).  It is minimized by ground states 
of two distinct kinds:
\begin{eqnarray}
s_m = s_{m+1}, & t_n = t_{n+1}, & 
\mbox{ Fig \ref{fig-cboard-vacua}(b) and (c),}
\label{chain-f}
\\
s_m = -s_{m+1}, & t_n = -t_{n+1}, & 
\mbox{ Fig \ref{fig-cboard-vacua}(d) and (e).}
\label{chain-af}
\end{eqnarray}  
Curiously, the two types of ground states (\ref{chain-f}) and
(\ref{chain-af}) are related to each other by the staggering
transformation (\ref{staggering}), rather than by any symmetry of the
lattice.  They remain degenerate even upon the inclusion of all two-
and four-chain interactions in Eq.~(\ref{4order}).  The origin of this
dynamical symmetry will be discussed in Sec.~\ref{sec-chains}.

\subsection{Long-range spin order}

It is natural to ask whether there is a long-range spin order in the
system.  Quantum effects turn out to be rather subtle in this case.
On the one hand, quantum fluctuations tend to destroy the long-range
order found on classical chains.  On the other, they create interchain
coupling and make the system two-dimensional thereby making spin order
more likely.  Which tendency wins?

To answer this question one can compute the expectation values of
local magnetization $\langle {\bf S}_{\bf r} \rangle$ in one of the
ground states [Fig.~\ref{fig-cboard-vacua}(b) through (e)].  The
classical value $(0,0,\pm S)$ is reduced by quantum fluctuations of
spins.  A na\"{\i}ve evaluation of this quantity at the first order in
$1/S$ gives a divergent negative correction suggesting that the order
is destroyed.  This, however, is an artefact of a low-order
approximation.  The magnon spectrum is given by the frequencies of {\em
classical} spin waves, which know nothing about the interchain
coupling.  As a resut, the magnon spectrum contains {\em lines} of
zero modes along the diagonal directions ($p_\xi=0, \pi$ and
$p_\eta=0,\pi$).  The abundance of soft modes leads to a destruction
of the long-range spin order.

At the next level of approximation, ${\cal O}(1/S^2)$, magnon
interactions modify the spin-wave spectrum lifting the zero modes to
finite frequencies (with the exception of isolated points in the
Brillouin zone).\cite{Chubukov91} The spin excitations become
two-dimensional and the infrared divergence of the correction to local
magnetization is removed.  Long-range order can survive.

Alternatively, the hardening of the spin-wave spectrum (specifically
of the zero modes) can be evaluated already at the order ${\cal
O}(1/S)$ by adding to the Heisenberg Hamiltonian a phenomenological 
biquadratic exchange term\cite{Henley87,Yildirim98}   
\begin{equation}
V_{\rm bi}= - K \sum_{\bf r, r'} 
\left( {\bf S}_{\bf r} \cdot {\bf S}_{\bf r'} \right)^2
\label{eq-bi}
\end{equation}
that couples spins at the intersections of chains.  The strength of this
interaction $K = {\cal O}(J_1^2/J_2 S^3)$ is chosen so as to mimick,
at the classical level, the energy of zero-point fluctuations.  The
latter is given, to the lowest order in $J_1$, by
Eq.~(\ref{eq-2-order}).  The spin-wave spectrum is then computed from 
the classical equations of motion for the spins.  

By performing a calculation along these lines (to be reported
elsewhere\cite{Starykh-unpub}) we find that indeed the zero modes
acquire energies of order $J_1 \sqrt{S}$.  The renormalized magnon
frequency vanishes at the points
\begin{equation}
(p_\xi, p_\eta) = (0,0) \mbox{ and } (\pi,\pi)
\end{equation}
only, as required by the Goldstone theorem.  The average local
magnetization reads
\begin{equation}
\langle S^z \rangle = S - 2\ln{\frac{J_2\sqrt{S}}{J_1}} 
+ \mbox{ regular terms.}
\label{eq-local-magn}   
\end{equation}
The correction to the classical value $S$ is a remnant of the
logarithmic divergence in one dimension that has been regularized by
an infrared cutoff $\omega_{\rm min} = {\cal O}(J_1 \sqrt{S})$ brought
about by the interchain coupling (\ref{eq-bi}).  The argument of the
logarithm is the ratio of the maximum magnon frequency $\omega_{\rm
max} = {\cal O}(J_2 S)$ to the infrared cutoff.

Taking this formula at face value we estimate that the N\'eel order is
present if the interchain exchange exceeds a critical value
\begin{equation}
J_{1c} = {\cal O}(J_2 \sqrt{S} e^{-S/2}). 
\label{eq-Jc}
\end{equation}
Note that the ordering is a truly collective effect since independent
spin chains possess no LRO.  This feature makes the order-by-disorder
problem rather different from its higher-dimensional analogues where
each individual unit (say, spin plane in a canonical example of two
interpenetrating square lattices) is ordered even in the absence of
frustrating inter-unit interactions.  Finally, Eq.~(\ref{eq-Jc})
indicates that at a large S there is a narrow region of ratios
$J_1/J_2$ without N\'eel order.  This result is in agreement with a
weak-coupling renormalization group analysis of the $S=1/2$ problem by
one of us\cite{Starykh02} who argued in favor of a gapless sliding
Luttinger liquid ground state in a wide interval $0 \leq J_1/J_2 \leq
0.8$.

\section{Crossed chains: $J_1<J_2$}
\label{sec-chains}

The four vacua [Fig.~\ref{fig-cboard-vacua}(b)--(e)] found in
the limit of weakly coupled chains, $J_1 \ll J_2$, remain the ground
states of the system for all finite ratios $J_1/J_2 < 1$.  To confirm
this, we have computed numerically spin-wave spectra of all collinear
classical ground states for a lattice $16 \times 16$ with periodic
boundary conditions.  Modulo the global O(3) spin symmetry, there are
$2^{16} = 65 536$ spin configurations to consider.  The energy of
zero-point motion (\ref{E-1}) is indeed minimized by the four N\'eel
states (\ref{chain-f}--\ref{chain-af}).  In fact, not only they remain
degenerate (with numerical accuracy): their spin-wave spectra are
identical.  That surely means that there is a hidden symmetry at work.

The observed fourfold degeneracy is caused by a special gauge-like
symmetry discovered by Henley.\cite{Henley-unpub} It exists whenever a
lattice can be split into corner-sharing units (tetrahedra on the
pyrochlore lattice, second-neighbor pairs in the present case) with
total zero spin in any ground state.  {\em Nonzero} eigenfrequencies
of such a system can be obtained by solving the equations of motions
for the transverse components of the total spins of these units $({\rm
Re}\sigma_{\alpha}, {\rm Im}\sigma_{\beta})$, which have the following
simple form:\cite{MCprl,MCprb}
\begin{equation}
i\, \dot{\sigma}_\alpha 
= \sum_{\beta \neq \alpha} J_{\alpha\beta} S_{\alpha\beta}\sigma_{\beta}.
\label{MC}
\end{equation}
Here $S_{\alpha\beta}$ is the ordered (longitudinal) component of the
spin shared by units $\alpha$ and $\beta$.  It can now be seen that
whenever two ground states are related by an Ising gauge transformation 
\begin{equation}
S'_{\alpha\beta} = \Lambda_{\alpha} S_{\alpha\beta} \Lambda^{-1}_{\beta},
\hskip 5mm
\Lambda_{\alpha} = \pm 1,
\label{gauge-link}
\end{equation}
their nonzero modes are also related,
\begin{equation}
\sigma'_{\alpha} = \Lambda_{\alpha} \sigma_{\alpha},
\label{gauge-site}
\end{equation}
and have identical frequency spectra.  Therefore gauge-equivalent ground
states have the same zero-point energy.

In the current context, sites $\alpha$ of the dual lattice are
``tetrahedra'' (squares with crossings) of the original checkerboard
lattice.  For equal exchanges $J_2 = J_1$, a unit contains four spins
whose total spin vanishes in a ground state.  For strong diagonal chains
($J_2>J_1$), the total spin must vanish on both diagonal bonds
separately, so that there are two units on every site of the dual
lattice: the diagonal links $\xi$ and $\eta$.  The resulting equations
for a collinear N\'eel state read
\begin{eqnarray}
i\dot{\mu}_{\bf r} &=& 
S_{\bf r, r+\hat{\xi}} 
(J_2\mu_{\bf r+\hat{\xi}} 
+ J_1\nu_{\bf r+\hat{\xi}}) 
\nonumber\\
&+&
S_{\bf r, r-\hat{\xi}} 
(J_2\mu_{\bf r-\hat{\xi}} 
+ J_1\nu_{\bf r-\hat{\xi}}) 
\nonumber\\
i\dot{\nu}_{\bf r} &=& 
S_{\bf r, r+\hat{\eta}} 
(J_1\mu_{\bf r+\hat{\eta}} 
+ J_2\nu_{\bf r+\hat{\eta}}) 
\nonumber\\
&+&
S_{\bf r, r-\hat{\eta}} 
(J_1\mu_{\bf r-\hat{\eta}} 
+ J_2\nu_{\bf r-\hat{\eta}}), 
\label{MC2}
\end{eqnarray}
where ${\bf r}$ are coordinates of a tetrahedron.  A spin labelled
${\bf S_{\bf rr'}}$ is shared by the tetrahedra located at ${\bf r}$
and ${\bf r'}$.  Finally, $\mu_{\bf r}$ and $\nu_{\bf r}$ are
transverse spin components of its units:
\begin{eqnarray}
{\bf S}_{m-\frac{1}{2},n} + {\bf S}_{m+\frac{1}{2},n} 
&=& ({\rm Re}\mu_{mn}, {\rm Im}\mu_{mn}, 0), 
\nonumber\\
{\bf S}_{m,n-\frac{1}{2}} + {\bf S}_{m,n+\frac{1}{2}} 
&=& ({\rm Re}\nu_{mn}, {\rm Im}\nu_{mn}, 0)
\end{eqnarray}
in the notations of Fig.~\ref{fig-cboard}.

The transformation (\ref{gauge-link}--\ref{gauge-site}) does not
actually reflect a local symmetry: applied to a single unit, $\alpha$,
it flips the spins $S_{\alpha\beta}$ shared by $\alpha$ with other
units $\beta$.  These other units acquire a nonzero total spin and
violate the ground-state condition.  Therefore the transformations
must be made on a number of dual sites (an infinite one for an
infinite lattice).  For $J_2>J_1$, entire diagonal chains of spins
must be flipped.  It can be checked that the two ground states shown
in Fig.~\ref{fig-cboard-vacua} (b) and (d) are related through such a
gauge transformation flipping spins on every other diagonal chain in
both directions.

In addition to breaking the spin O(3) symmetry, the 4 ground states
also violate the spatial symmetry of the checkerboard lattice.  This
leads to interesting consequences.  Although the spin symmetry must be
restored at any finite temperature (the Mermin--Wagner theorem), the
discrete lattice symmetries need not.  In such a case, the
low-temperature phase can have a long-range spin-Peierls (bond) order.
An example of such behavior was discovered by Chandra 
{\em et al.}\cite{Chandra} 
for the Heisenberg antiferromagnet on the square lattice
with large second-neighbor coupling.  

To see the pattern of bond order in the proposed spin-Peierls states
one can look at the bond averages $\langle {\bf S}_i \cdot {\bf S}_j
\rangle$ in the ground states of Fig.~\ref{fig-cboard-vacua} (b)--(e).
We have shown the frustrated bonds (those with parallel spins) in
color depending on the bond orientation: red for vertical and green
for horizontal ones.  The lower part shows the dual lattice of
``tetrahedra'' each painted in the corresponding color.  Using this
language, the ground states can be described in a simple manner: {\em
second} neighbors on the dual lattice have the same color.  In fact,
if the dual sublattice is divided into two sublattices, each
sublattice exhibits ``ferromagnetic'' Ising order---in terms of these
color variables---independently of the other sublattice.  Hence a
fourfold degeneracy mentioned above.

An ordered state thus can be completely characterized by a composite
order parameter ${\bm Z}_2 \times {\bm Z}_2 \times {\bm S}_2$: two
independent Ising variables and a N\'eel vector.  The Ising order
parameter,
\begin{equation}
f_2 = \langle ({\bf S}_{\bf r,r+\hat{\xi}} 
- {\bf S}_{\bf r,r-\hat{\xi}}) \cdot
({\bf S}_{\bf r,r+\hat{\eta}} 
- {\bf S}_{\bf r,r-\hat{\eta}} 
) \rangle, 
\label{f2}
\end{equation}
has a counterpart in the frustrated antiferromagnet on the square
lattice with a second-neighbor coupling.\cite{Chandra} However, we
will find a richer structure of ground states because our order
parameter is, in fact, a component of a doublet ${\bf f} = (f_1,
f_2)$, defined for every tetrahedron: 
\begin{eqnarray}
f_1 & = & \frac{\langle({\bf S}_1 + {\bf S}_2)\!\cdot\!({\bf S}_3 +
{\bf S}_4) - 2({\bf S}_1 \cdot {\bf S}_2 + {\bf S}_3 \cdot {\bf S}_4)
\rangle} {\sqrt{12}}, \nonumber \\ f_2 & = & \frac{\langle({\bf S}_1 -
{\bf S}_2) \cdot ({\bf S}_3 - {\bf S}_4)\rangle}{2}.
\label{eq:f-def}
\end{eqnarray}
[Fig.~\ref{fig-colors}[(a)].  The other component, $f_1$, comes into play
when the vertical and horizontal bonds become (nearly) equivalent to
the diagonal ones, a situation encountered on the three-dimensional
pyrochlore lattice.\cite{Tchernyshyov02a} 

\section{Planar pyrochlore: $J_1 = J_2$}
\label{sec-cboard}

This is a point with a very large {\em classical}
degeneracy.\cite{MCprl,MCprb} Only the total spin of a ``tetrahedron''
must vanish in a ground state, but not necessarily the spins of
second-neighbor pairs separately.  To the next order, ${\cal O}(1/S)$,
numerical comparison of zero-point fluctuation energies in collinear
N\'eel states still reveals a large degeneracy---much larger than in
the previously discussed case $J_2>J_1$.  The ground states of
Fig.~\ref{fig-cboard-vacua} (b)--(e) become degenerate with the
N\'eel state of the simple square lattice
[Fig.~\ref{fig-cboard-vacua}(a)] and many others numbering $2^{L}$ in
total.  Apparently this multitude of degenerate ground states kills
the long-range N\'eel order.  On the other hand, it will be seen that
the bond order survives.

To proceed, we present an explicit construction of all collinear 
ground states degenerate at the $1/S$ level, and identify a short-range
interaction that selects these ground states.  It turns out that the 
most economical description of these states is obtained in terms of
the bond---rather than spin---variables.  

\subsection{Gauge-equivalent collinear states}

For equal exchanges, Eq.~(\ref{MC}) holds for quartets of spins on
``tetrahedra'' of the checkerboard lattice (squares with crossings).
As far as collinear N\'eel states are concerned, there are now three
distinct possibilities: parallel spins can be found on vertical,
horizontal, and now also diagonal bonds, which we encode,
respectively, as red, green, and blue states of a tetrahedron.  In the
N\'eel state of the simple square lattice
[Fig.~\ref{fig-cboard-vacua}(a)] all diagonal bonds have parallel
spins, so that this state is uniformly blue.  After casting Henley's
gauge principle in bond language, we will readily reproduce all
collinear N\'eel states degenerate with the blue vacuum.  The four
ground states found in the $J_2>J_1$ case [Fig.~\ref{fig-cboard-vacua}
(b)--(e)] are among these.

As before, a gauge transformation on a tetrahedron involves flipping
all its spins.  Parallel spins are found on the same bonds before and
after the transformation, therefore the color of that tetrahedron
remains unchanged; it is the colors of its neighbors that are
affected.  Therefore, gauge transformations can be done {\em
separately and independently} on the two sublattices of
tetrahedra.

Flipping the four spins $S_{\alpha\beta}$ on tetrahedron $\alpha$ from
sublattice A takes four adjacent tetrahedra $\beta \in$ B out of the
ground state.  To fix this problem, for each $\beta$ we must perform
at least one more gauge transformation on one of {\em its} neighbors
$\gamma \in$ A.  Here are the rules for gauge transformations
performed on sublattice A:

\begin{quote}
(a) For every tetrahedron $\beta \in$ B, the number of gauge
transformations $\Lambda_\alpha = -1$ on adjacent tetrahedra $\alpha
\in$ A can be 0, 2, or 4.  

(b) If this number is 0 or 4, $\beta$ remains blue.

(c) For 2 gauge transformations, the two $\alpha$ {\em cannot} be on opposite
sides of $\beta$ (e.g., northeast and southwest).

(d) If both $\alpha$ are north of $\beta$ (or both are south of $\beta$),
$\beta$ becomes red.  If both $\alpha$ are east (west) of $\beta$, it
turns green.
\end{quote}

\begin{figure}
\includegraphics[width=\columnwidth]{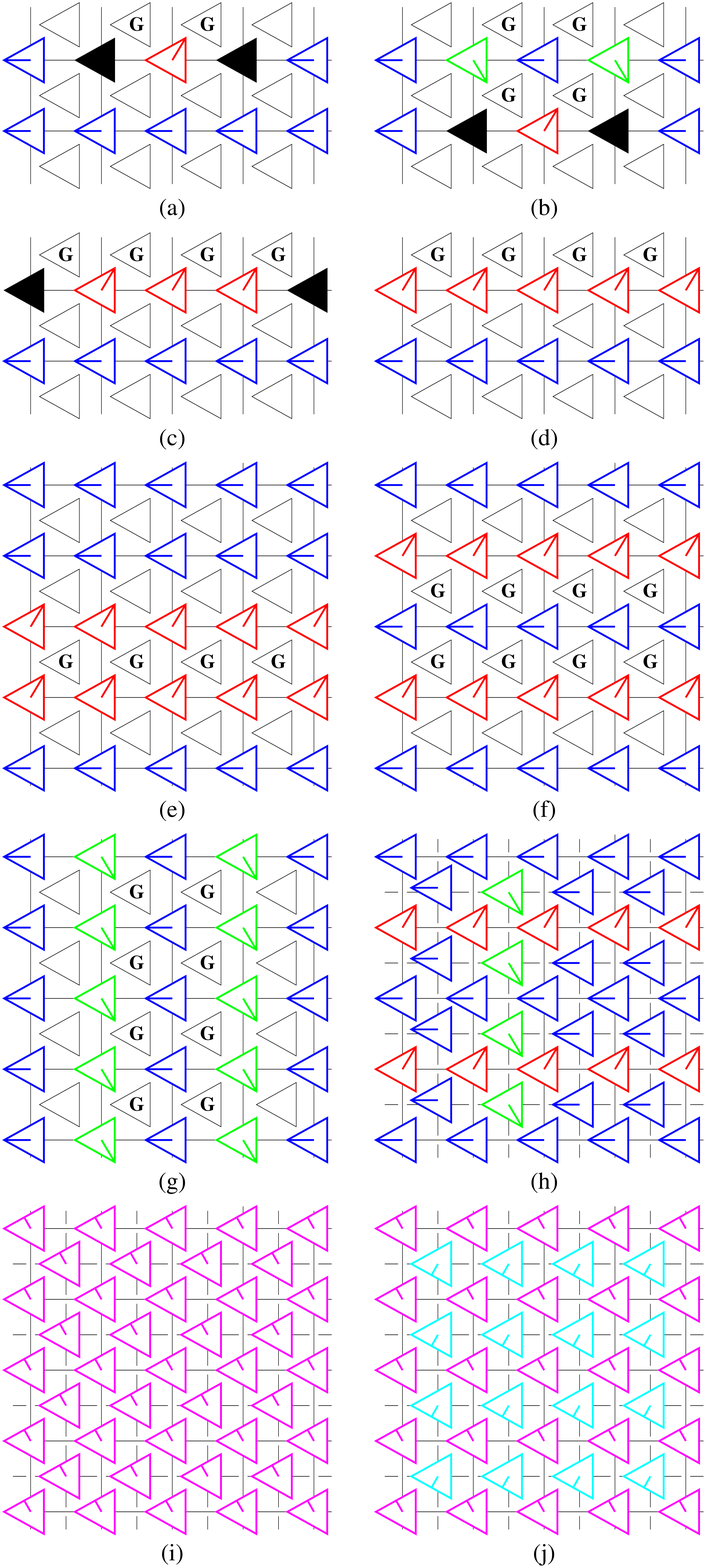} 
\caption{Construction of ground states at the order $1/S$ for
$J_1=J_2$.  The secondary color magenta is an equal-part mixture of
red and blue; similarly, cyan is an equally-weighted average of green
and blue.  The letter G marks the locations of ${\bm Z}_2$ gauge
transformations. Filled triangles represent tetrahedra not in a
ground state.}
\label{fig-constr}
\end{figure}

By using these rules, we can now construct an arbitrary ground state
starting with the blue one.  As the ground state is a direct product
of independent ground states on sublattices A and B, we will construct
a ground state of sublattice B by making gauge transformations on
sublattice A.  Suppose, for definiteness, that there is a red
tetrahedron on sublattice B and that the two gauge transformations
were made north of it [Fig.~\ref{fig-constr}(a)].  If no other gauge
transformations were made, this state would violate Rule (a): the two
B tetrahedra shown in black are not in their ground states.
Additional gauge transformations cannot be made around the original
red site, for it will become blue [Fig.~\ref{fig-constr}(b)], contrary
to the initial assumption.  The only remaining possibility is shown in
Fig.~\ref{fig-constr}(c): additional gauge transformations are made on
the same horizontal line.  Continuing the process we find a line of
red tetrahedra extending over the entire sublattice B in the
horizontal direction [Fig.~\ref{fig-constr}(d)].  Thus a generic
ground state of one sublattice consists of horizontal red and blue
stripes of arbitrary widths [Fig.~\ref{fig-constr}(e,f)] or of
vertical green and blue stripes [Fig.~\ref{fig-constr}(g)].

The process is then repeated with the roles of the sublattices
reversed: sublattice A is colored via gauge transformations on
sublattice B (whose colors are unchanged).  A sample ground state is
shown in Fig.~\ref{fig-constr}(h).  

Now it is easy to count the number of degenerate ground states on an
$L \times L$ checkerboard lattice with periodic boundary conditions.
On a single sublattice of tetrahedra, there are $2 \times 2^{L/2 - 1}$
red-and-blue ground states: the exponential reflects the number of
ways to place horizontal domain walls separating red and blue stripes;
the prefactor accounts for a duplicate set of states with red and blue
domains exchanged.  In addition, there is an equal number of
green-and-blue states bringing the total to $2^{L/2+1} - 1$ per
sublattice.  (The blue state has been counted twice.)  The total
degeneracy of the ground state (including both sublattices of
tetrahedra) is therefore of order $2^L$.  

\subsection{Long-range bond order}

Is there a spontaneously broken symmetry?  There's definitely no
long-range {\em vector} order.  It is easy to see that two spins
located on the same diagonal can be parallel and antiparallel with an
equal probability: they are parallel if there is an even number of
nonblue tetrahedra in between, antiparallel if the number is odd.  By
construction (Fig.~\ref{fig-constr}), the probabilities of these
outcomes are equal and $\langle {\bf S}_{\bf r} \cdot {\bf S}_{\bf
r'}\rangle = 0$ for these two spins: there is not even a short-range
order.  The argument can be extended to (almost) any other direction.
(Exceptions are the vertical and horizontal directions: there are
long-range correlations of spins along the horizontal or vertical
lines of bonds in the cyan and magenta states.)

The only remnant of the spin order is the collinearity of spins:
$({\bf S}_{\bf r} \cdot {\bf S}_{\bf r'})^2 = S^4$ in any ground state
for any pair of spins.  Thus one can conclude that the ground state is
a spin nematic whose order parameter is a {\em director} (${\bm
S}^2/{\bm Z}_2$).

In addition, the system has long-range {\em bond} order.  Indeed, if
one tetrahedron on sublattice A is colored red, there are no green
tetrahedra anywhere on the same sublattice---and vice versa.  Thus the
symmetry between red and green colors (the symmetry between vertical
and horizontal bonds) is spontaneously broken. The average color of a
given tetrahedron is magenta (red and blue stripes) or cyan (green and
blue).  Each sublattice finds itself in the cyan or magenta phase,
Figs.  \ref{fig-constr}(i,j).  In the cyan phase, the vertical
(``red'') bonds feature antiparallel spins; in the magenta phase,
antiparallel spins are found on horizontal (``green'') bonds.  The
colors of the sublattices are {\em independent} at order $1/S$.

This argument can be made more precise by turning to the bond order
parameter $f_2$ (\ref{eq:f-def}).  As shown in
Fig.~\ref{fig-colors}(a), $f_2$ is positive, zero, and negative in the
red, blue, and green states, respectively.  Averaging over the entire
ground state manifold gives $\langle f_2({\bf r}) \rangle = 0$ for any
tetrahedron.  The reasoning put forward in the previous paragraph
suggests that $\langle f_2({\bf r}) f_2({\bf r'}) \rangle = S^4/2 > 0$
even as $|{\bf r - r'}| \to \infty$ (as long as the tetrahedra ${\bf
r}$ and ${\bf r'}$ reside on the same sublattice).  Thus a long-range
bond order is present.\cite{Chaikin}  

The long-range order is thus similar to the case $J_1<J_2$, with one
exception: the N\'eel vector in the composite order parameter is
replaced with a director that defines a collinearity axis.  The order
parameter now has the structure ${\bm Z}_2 \times {\bm Z}_2 \times
{\bm S}^2 / {\bm Z}_2$: two Ising order parameters in addition to a
spin nematic.  The spin-nematic order breaks a continuous rotational
symmetry and therefore will be lost at any finite temperature.  The
remaining Ising orders ${\bm Z}_2 \times {\bm Z}_2$ are expected to
survive up to a finite temperature $\mathcal O(JS)$.

\subsection{Effective spin interaction}
\label{sec-effective}

One may wonder what kind of a Hamiltonian gives rise to a strongly
degenerate set of ground states described above.  In fact, it can be
derived following Henley's method.\cite{Henley-unpub} The quantum
correction of order $1/S$ to the ground state energy (\ref{E-1}) is
obtained by rewriting Eq.~(\ref{MC}) for the eigenmodes:
\begin{eqnarray}
\hbar^2 \omega^2 \sigma_\alpha 
&=& J^2 \sum_{\beta(\alpha)} \sum_{\gamma(\beta)} 
S_{\alpha\beta} S_{\beta\gamma} \sigma_\gamma
\nonumber\\
&=& 4 J^2 S^2 \sigma_\alpha 
+ J^2 \sum_{\beta(\alpha)} \sum_{\gamma(\beta)\neq\alpha} 
S_{\alpha\beta} S_{\beta\gamma} \sigma_\gamma.  
\end{eqnarray}
The notation $\beta(\alpha)$ indicates that tetrahedron $\beta$ is a
neighbor of tetrahedron $\alpha$ (they share spin $S_{\alpha\beta}$).
Following Henley, we introduce an adjacency matrix, 
\begin{equation}
T_{\alpha\gamma} = - \frac{1}{2S^{2}} \sum_{\beta(\alpha,\gamma)} 
S_{\alpha\beta} S_{\beta\gamma},  
\end{equation}
whose matrix elements are nonzero when tetrahedra $\alpha$ and
$\gamma$ have a common neighbor $\beta$.  The sum over nonzero
eigenfrequences can now be expressed in terms of the matrix $T$:
\begin{equation}
\sum_{n} \frac{\hbar |\omega_n|}{2} = JS\, {\rm Tr} \sqrt{1-T} 
= JS\, {\rm Tr} \left(1 - \frac{T}{2} - \frac{T^2}{8} - \ldots\right).
\label{E-1-T}
\end{equation}
This expansion converges rather slowly (eigenvalues of $T$ extend all
the way up to 1) and cannot be used for quantitative purposes.
Nevertheless, it provides correct qualitative answers, as will be seen
shortly.

\begin{figure}
\includegraphics[width=\columnwidth]{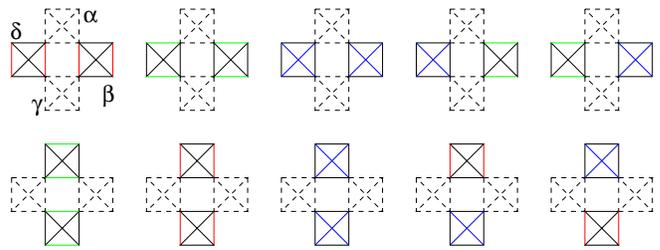} 
\caption{Ground states of Henley's four-spin interaction,
Eq.~(\ref{four-spin}).  Frustrated bonds are shown in color.
Parametrization in terms of bond (rather than spin) variables
decouples tetrahedra of one sublattice ($\alpha, \gamma$) from those
of the other ($\beta, \delta$).  }
\label{fig-henley}
\end{figure}

Because $T_{\alpha\alpha} = 0$, the first spin-dependent correction
comes from the second-order term,
\begin{equation}
-\frac{JS}{4} {\rm Tr} T^2 
= -3JSL^2 -\frac{J}{4S^3} 
\sum_{\alpha\beta\gamma\delta} 
S_{\alpha\beta} S_{\beta\gamma} S_{\gamma\delta} S_{\delta\alpha}.
\label{four-spin}
\end{equation}
(In the last sum, all tetrahedra are distinct.)  This spin-dependent
term can be considered as a four-body interaction of spins or as a
two-body interaction of bond variables.  Indeed, the interaction 
$-S_{\alpha\beta} S_{\beta\gamma} S_{\gamma\delta} S_{\delta\alpha}$ 
is minimized when the number of down spins is even, which can be cast 
in bond language as 
\begin{equation}
S_{\alpha\beta} S_{\beta\gamma} 
= S_{\gamma\delta} S_{\delta\alpha} = \pm S^2, 
\end{equation}
i.e., when the {\em vertical} bonds of tetrahedra $\beta$ and $\delta$
simultaneously have parallel or antiparallel spins.  All such bond
configurations are shown in the top row of Fig.~\ref{fig-henley}. 
Alternatively, this 4-spin term can be considered as a bond interaction
between tetrahedra $\alpha$ and $\gamma$, which is minimized when 
\begin{equation}
S_{\delta\alpha} S_{\alpha\beta} 
= S_{\beta\gamma} S_{\gamma\delta} = \pm S^2, 
\end{equation}
i.e., when their {\em horizontal} bonds simultaneosuly have parallel
or antiparallel spins.  These ground states are shown in the bottom
part of Fig.~\ref{fig-henley}.  

Note that the bond interaction generated by the term $-{\rm Tr} T^2$
is between tetrahedra of the same sublattice.  Higher-order spin loops
$-{\rm Tr} T^n$ can also be represented as an interaction between $n$
tetrahedra of the same sublattice.  In terms of bond variables, there
is no coupling between different sublattices at the order $1/S$.

Before deriving the three-body interaction, it makes sense to check
for the ground states of the two-body bond Hamiltonian that we have
just obtained: the energy is $-1$ for states shown in
Fig.~\ref{fig-henley} and $+1$ for the remaining states.  More
concisely,
\begin{quote}
For tetrahedra on the same sublattice, nearest neighbors in the
horizontal direction need to be both red or neither red; in the
vertical direction both green or neither green.
\end{quote}
It is easy to see that this rule gives precisely all the ground states
found in the beginning of this Section by means of Henley's gauge
argument---and no other states.  

It is quite remarkable that a crude two-body approximation, 
\begin{equation}
{\rm Tr}\sqrt{1-T} \approx {\rm Tr}(1 - T/2 - T^2/8),
\end{equation}
correctly reproduces all the ground states.  This fact seems to
indicate that the neglected $n$-body interactions (which are by no
means small) can be expressed in terms of the two-body potentials
identified above.  If this is the case, the many-body interactions
shift all ground states by the same amount without breaking their
degeneracy.  (We have checked that the 3 and 4-body interactions are
indeed reducible.)

\subsection{Effective bond interaction}

The four-spin interaction (\ref{four-spin}) can also be written in terms 
of the bond variables (\ref{eq:f-def}).  Adjacent tetrahedra $\alpha$ 
and $\gamma$ of the same sublattice (Fig.~\ref{fig-henley}) 
interact with energy
\begin{equation}
E_{\alpha\gamma} = -\frac{JS}{8}
\left( \frac{1}{3} 
+ \frac{{\bf f}_\alpha \cdot \hat{\bf n}_{\alpha\gamma}}{S^2\sqrt{3}} \right)
\left( \frac{1}{3} 
+ \frac{{\bf f}_\gamma \cdot \hat{\bf n}_{\alpha\gamma}}{S^2\sqrt{3}} \right).
\label{two-f}
\end{equation}
The unit vector $\hat{\bf n}_{\alpha\gamma}$ points in the red
direction [Fig.~\ref{fig-colors}(a)] for neighbors along $\hat{\bf x}$
and in the green direction for neighbors along $\hat{\bf y}$.  The
one-body piece can be viewed as a ``magnetic'' field for the
two-component ``spin'' ${\bf f}_\alpha$ that points in the cyan or
magenta direction.  Once we sum over all neighbors of a given
tetrahedron, the resulting ``magnetic field'' has the blue color and
the magnitude $J/4S^3\sqrt{3}$.  The two-body potential couples
ferromagnetically the red or green components of neighboring
``spins'' depending on the direction.

It is worth noting that a very similar interaction has been obtained
for the quantum case $S=1/2$ by
Tsunetsugu.\cite{Tsunetsugu01,Tsunetsugu02} We will return to this
point later in Sec.~\ref{sec-quantum}.

\begin{figure}
\includegraphics[width=0.98\columnwidth]{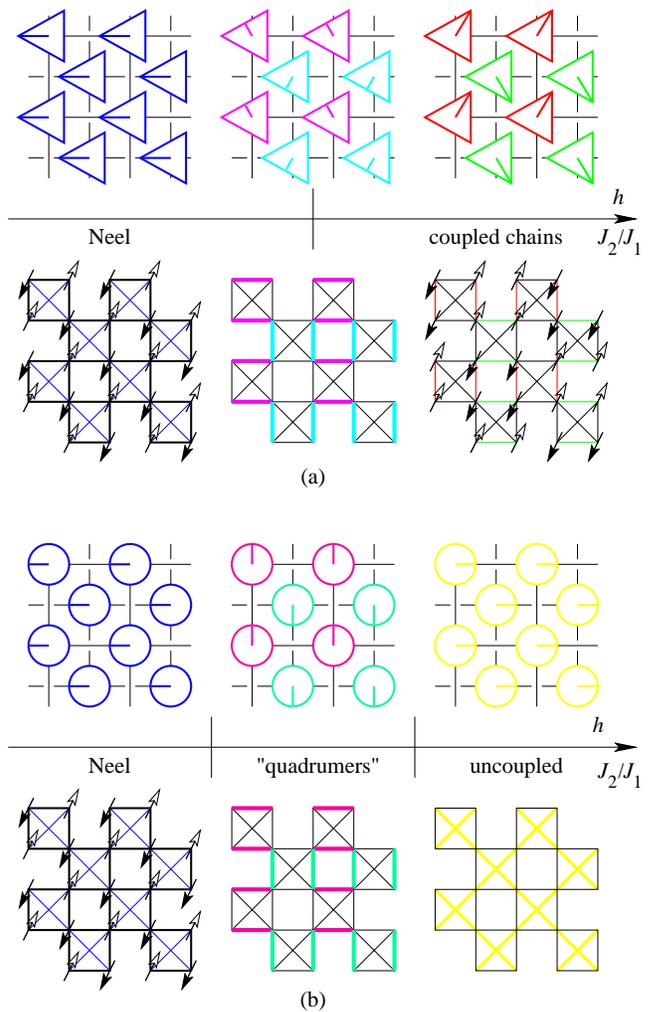} 
\caption{(a) Ground states computed to order $1/S$ as a function of
$J_2/J_1$.  The two sublattices of tetrahedra are colored
independently of each other.  (b) Ground states for $S=1/2$
(Refs.~\onlinecite{Starykh02,Fouet01,Sindzingre02,Berg02}).  The secondary
colors (cyan, magenta, and yellow) encode the location of satisfied
bonds (those with antiparallel spins).  They are the opposites of the
primary-color states (respectively, red, green, and blue).  }
\label{fig-phases}
\end{figure}

\section{Large $S$: a brief summary}
\label{sec-summary}

Our study of the ground state of the Heisenberg checkerboard
antiferromagnet to order $1/S$ establishes the existence of a
long-range bond order in this system.  The bond order breaks a mirror
symmetry of the lattice exchanging the vertical and horizontal
directions (the green and red colors in our notation).  The discrete
(${\bm Z}_2$) nature of the broken symmetry assures survival of the
long-range bond order to finite temperatures.  It is instructive to
trace the evolution of the ground state as the ratio of the first and
second-neighbor exchange couplings varies.

For $J_2/J_1<1$, the ground state is unique (up to a global rotation of
all spins): it is the N\'eel state of the simple square lattice ($J_2
= 0$).  Although the global O(3) spin symmetry is broken at $T=0$, 
it is restored at any finite temperature.  The ${\bm Z}_2$ symmetry
is manifest.  

For $J_2/J_1>1$, we find a fourfold degeneracy as each sublattice
independently chooses one of the two ground states with collinear
spins.  In the red state, {\em parallel} spins are found on vertical
bonds (and antiparallel spins on all other bonds).  In the green
state, parallel spins are found on horizontal bonds
(Fig.~\ref{fig-cboard-vacua}).

The point $J_2/J_1=1$ is special: there are ${\cal O}(2^L)$ degenerate
ground states on an $L\times L$ lattice to order $1/S$.  There is no
N\'eel order even at zero temperature.  Still, long-range bond order
is present: if a single tetrahedron is put in the red state, there are
no green tetrahedra on the same sublattice.  By averaging over all
red-and-blue states (Fig.~\ref{fig-constr}) one obtains a magenta
phase in which {\em antiparallel} spins, $\langle {\bf S}_i \cdot {\bf
S}_j \rangle = -S^2$, are found on horizontal bonds.  (For the rest of
first and second neighbors, $\langle {\bf S}_i \cdot {\bf S}_j \rangle
= 0$.)  The cyan phase, where $\langle {\bf S}_i \cdot {\bf S}_j
\rangle = -S^2$ on vertical bonds and 0 on the rest, is obtained by
averaging over the green-and-blue states.

The independent ordering of the two sublattices of tetrahedra 
is probably 
accidental: the quantum correction of order $1/S$ (\ref{E-1-T})
couples bond variables of the same sublattice.  Higher-order
corrections may well produce intersublattice couplings locking the
sublattice order parameters.

As we have shown for $J_2/J_1=1$, parametrization of the four-spin
interaction (\ref{four-spin}) in color terms gives a Potts-like model
with $q=3$ states and direction-dependent interactions
(Fig.~\ref{fig-henley}).  Small deviations of the difference $J_2/J_1$
from 1 amount to adding a magnetic field $h$ promoting or suppressing
the blue state.  For $J_2/J_1<1$ (blue field), it is advantageous to
place {\em parallel} spins on the weaker diagonal bonds making the
entire lattice blue.  For $J_2/J_1>1$, the field points in the
opposite---yellow---direction suppressing the blue state and forcing
the system to choose between the red and green states.  Lastly,
$J_2/J_1 = \infty$ is the regime of decoupled diagonal chains.  The
phase diagram is shown schematically in Fig.~\ref{fig-phases}(a).

\section{Quantum limit: $S = 1/2$}
\label{sec-quantum}

It is interesting to compare our large-$S$ answers to the numerical
results for $S=1/2$ obtained
recently.\cite{Fouet01,Sindzingre02,Berg02} For $J_1 = J_2$ they have
found a bond-ordered state in which the probability of finding a spin
singlet is enhanced on half of the squares without crossings, the
``quadrumer'' phase in Fig.~\ref{fig-phases}(b). A similar plaquette
state has appeared in the analyses of the quantum dimer model for the
planar pyrochlore\cite{Moessner01} and the Shastry-Sutherland
lattice.\cite{Chung} In our terminology, this plaquette state
corresponds to the magenta-cyan vacuum [Fig.~\ref{fig-phases}(a)]: the
spin correlations $\langle {\bf S}_i \cdot {\bf S}_j \rangle$ are more
negative on vertical bonds for one sublattice of tetrahedra and on
horizontal bonds for the other. The $S=1/2$ ground state quantum
wavefunction, however, cannot be expressed as a simple product of
single tetrahedron configurations.

Although the large-$S$ and $S=1/2$ answers are the same at equal
couplings, there are also important differences that become apparent
when we compare the results for $J_1 \neq J_2$.  In the large-$S$
phase diagram [Fig.~\ref{fig-phases}(a)], the magenta-cyan state is a
single point sandwiched between the blue and red-and-green phases.
For $S=1/2$, it is found in a finite range of ratios $J_2/J_1$ around
1.  The phase diagram for $S=1/2$ inferred from the works of Lhuillier
{\em et al.} is shown in Fig.~\ref{fig-phases}(b).

The $S=1/2$ case does not lend itself to a straightforward analytical
treatment.  Nevertheless, we have found a useful phenomenological
approach that sheds some light onto its phase diagram.  The bond
variables ${\bf f}$, which we have introduced previously for classical
spins (\ref{eq:f-def}), can be defined in the same way for any spin
value $S$.  The ground states of an isolated tetrahedron with a
Heisenberg interaction between its four spins are $2S+1$ degenerate
singlets.  The operators $\hat{\bf f} = (\hat{f}_1, \hat{f}_2)$,
\begin{eqnarray}
\hat{f}_1 & = & \frac{({\bf S}_1 + {\bf S}_2)\!\cdot\!({\bf S}_3 +
{\bf S}_4) - 2({\bf S}_1 \cdot {\bf S}_2 + {\bf S}_3 \cdot {\bf S}_4)
} {\sqrt{12}}, \nonumber 
\\ 
\hat{f}_2 & = & \frac{({\bf S}_1 -
{\bf S}_2) \cdot ({\bf S}_3 - {\bf S}_4)}{2}.
\label{eq:f-quant}
\end{eqnarray}
leave this manifold of states invariant.  In this Hilbert space they 
act as Hermitian matrices $(2S+1)\times(2S+1)$.\cite{Tchernyshyov02b}  
In particular, for $S=1/2$ they are proportional to the Pauli matrices.
One can choose a basis in which 
\begin{equation}
\hat{f}_1 = \frac{\sqrt{3}}{2}\tau_x, 
\hskip 5mm
\hat{f}_2 = \frac{\sqrt{3}}{2}\tau_z.
\label{eq-f-half}
\end{equation}

\begin{figure}
\includegraphics[width=0.98\columnwidth]{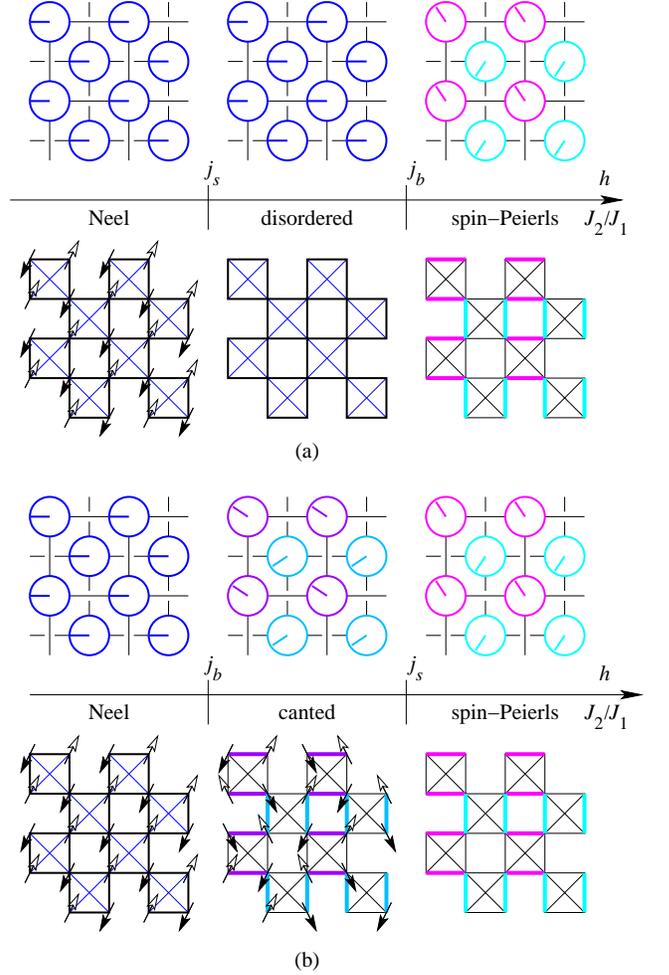} 
\caption{Transitions between the N\'eel and ``quadrumer'' states. 
(a) Spin order melts at $J_2/J_1 = j_s$, bond order appears at 
$J_2/J_1 = j_b > j_s$. 
(b) Spin order becomes noncollinear at $J_2/J_1 = j_b$, then disappears
at $J_2/J_1 = j_s > j_b$.  
}
\label{fig-phases-detail}
\end{figure}

If the interaction between the bond variables on different tetrahedra
$\alpha$ and $\beta$ were of the pure Potts form, it would be
proportional to the scalar product ${\bf f}_\alpha \!\cdot\! {\bf
f}_\beta$.  In the $S=1/2$ case, 
\begin{equation}
{\bf f}_\alpha \!\cdot\! {\bf f}_\beta
= \frac{3}{4}(\tau_{\alpha x} \tau_{\beta x} 
+ \tau_{\alpha z} \tau_{\beta z}). 
\end{equation}
However, as we have seen in Sec.~\ref{sec-effective}, the interaction
has a more complicated form.  It is direction-dependent and is
asymmetric in the Potts flavors: red and green are different from the
blue even when $J_1 = J_2$.  The crudest way to reflect this asymmetry
is to write an interaction of the form $K_1 f_{\alpha 1} f_{\beta 1} +
K_2 f_{\alpha 2} f_{\beta 2}$ with $K_1 \neq K_2$.  In addition, there
can be a ``magnetic field'' coupling $-h f_{\alpha 1}$ that selects
the blue states (for $J_2 \ll J_1$) or the red and green states ($J_2
\gg J_1$) in the classical case.
\begin{equation}
H = \sum_{\langle\alpha\beta\rangle}
[K_{xx}(R_{\alpha\beta}) \tau_{\alpha x} \tau_{\beta x}
+ K_{zz}(R_{\alpha\beta}) \tau_{\alpha z} \tau_{\beta z}]
- h \sum_{\alpha} \tau_{\alpha x}.
\end{equation}
Here we assume for simplicity that the potentials $K_{xx}$ and
$K_{zz}$ depend on the distance between tetrahedra, but not on the
direction.  Our findings in the previous section indicate that the
second-neighbor coupling is ferromagnetic ($K<0$).  Numerical results
of the Fouet {\em et al.}\cite{Fouet01} suggest that the
nearest-neighbor interaction is antiferromagnetic ($K>0$).  The fact
that the Ising order parameter is always related to the component
$f_2$ points to the dominance of the $K_{zz}$ coupling over $K_{xx}$.
In this case we can neglect the $K_{xx}$ terms altogether without 
changing the critical properties in any significant way: 
\begin{equation}
H = \sum_{\langle\alpha\beta\rangle}
K(R_{\alpha\beta}) \tau_{\alpha z} \tau_{\beta z}
- h \sum_{\alpha} \tau_{\alpha x}.
\label{eq-H-half}
\end{equation}
This is the Hamiltonian of an Ising antiferromagnet in a transverse
magnetic field, whose properties are well known.  At $h=0$, it is a
classical system with long-range order: $\langle\tau_z\rangle = +1$ on
one sublattice and $-1$ on the other.  (The tetrahedra form a square
lattice, so that the interactions are not frustrated.)  The N\'eel
phase is preserved in a finite range of weak magnetic fields $|h|
\lesssim K$.  A strong ``magnetic field'' $h \gg K$ induces fast
quantum fluctuations that kill the N\'eel order even at zero
temperature, so that $\langle\tau_z\rangle = 0$.  The zero-temperature
phase diagram is shown in Fig.~\ref{fig-phases} (b).  The system
(\ref{eq-H-half}) has an ``antiferromagnetic'' phase for $|h| < h_c$,
which corresponds to the ``quadrumer'' state of Fouet {\em et
al.}\cite{Fouet01} Quantum critical points at $h = \pm h_c$ separate
this phase from the quantum disordered regions with no bond
order---the yellow and blue states in which the diagonal bonds are
strong ($\langle \tau_x \rangle > 0$) and weak $\langle \tau_x \rangle
< 0$), respectively.

Let us pause for a moment and take a critical look at this
phenomenology.  From the viewpoint of bond variables, at some value
$J_2/J_1 = j_c \approx 1$, which corresponds to $h=0$, conditions are
ideal for a spin-Peierls state: quantum fluctuations induced by the
``transverse field'' $h \propto J_2/J_1 - j_c$ are absent, so that bond
order is robust.  Going away from that point increases quantum
fluctuations of $f_2 \propto \tau_z$ and reduces the bond order
parameter until the bond order completely melts at a quantum critical
point.  If the critical behavior is adequately described by the Ising
model, the energy gap vanishes at the critical point only.  Both
phases---bond-ordered and bond-disordered---are gapped.

This may or may not be the case.  Consider in more detail the
transition from the N\'eel phase of the simple square lattice
($J_2/J_1 = 0$) to the spin-Peierls phase at $J_2/J_1 \approx 1$.
There are three distinct possibilities: (1) As $J_2/J_1$ increases,
first the spin order melts at $J_2/J_1 = j_s$, then a bond order
appears at $J_2/J_1 = j_b > j_s$, see Fig.~\ref{fig-phases-detail}
(a).  The intermediate phase has neither spin, nor bond order.  The
critical behavior near $J_2/J_1 = j_b$ is adequately described by our
bond phenomenology.  (2) Magnetic order persists into the spin-Peierls
phase.  The spin and bond orders coexist\cite{Sachdev02} in the range
$j_b < J_2/J_1 < j_s$.  In this case, the emergence of bond order at
$J_2/J_1 = j_b$ is more adequately described as a spin canting
transition, see Fig.~\ref{fig-phases-detail} (b).  The system is
gapless on both sides of this quantum phase transition.  (3) The spin
order disappears simultaneously with the onset of the bond order at
$J_2/J_1 = j_s = j_b$, as shown in Fig.~\ref{fig-phases} (b).  One
phase is gapped, the other is gapless.

As is the usual limitation of exact diagonalisations, the numerical
data of Sindzingre {\em et al.}\cite{Sindzingre02} were obtainable
only for moderately small lattices (up to $6\times6$).  On the basis
of their data we are unable to tell whether or not the spin and bond
orders overlap.  We feel, however, that all three possibilities can be
realised as a matter of principle, so the Ising phenomenology may be
useful.

\section{Conclusion}

In this paper, we have studied the ground state of a Heisenberg
antiferromagnet with large spins $S$ on the checkerboard lattice, also
known as the planar pyrochlore.  To zeroth order in $1/S$---the
classical approximation---the magnet has an extremely large,
continuous degeneracy of the ground state.  In the next order in
$1/S$, this accidental degeneracy is partially lifted by quantum
fluctuations.  The main achievement of this work is a complete
characterization of the ground-state properties of this magnet to
order $1/S$.

The ground states with the lowest energy of zero-point motion are
found among the classical vacua with collinear spins.  By using a
special dynamical symmetry discovered by Henley,\cite{Henley-unpub} we
have explicitly constructed all of these ground states and shown that
their number is of order $2^L$ in a lattice $L\times L$.  We have
shown that there is no long-range N\'eel order: in the ensemble of
these ground states spin correlations $\langle {\bf S}_i \cdot {\bf
S}_j\rangle$ vanish beyond nearest neighbors.  However, there is a
long-range {\em bond} order: ``tetrahedra'' of the checkerboard
lattice spontaneously pick up one of the two states, in which the
nearest-neighbor correlations $\langle {\bf S}_i \!\cdot\! {\bf
S}_j\rangle$ are uniformly $-S^2$ for vertical bonds and 0 for
horizontal ones---or vice versa.  The bond order breaks the rotational
symmetry of the lattice.  

More precisely, the bond order parameter is ${\bm Z}_2 \!\times\!  {\bm
Z}_2$: the two sublattices of tetrahedra order {\em independently} of
each other, which is reminiscent of the Ashkin--Teller model.  The
discrete character of the broken symmetry indicates that the bond
order will likely survive at low temperatures.  The critical
properties of the thermal transition between the bond-ordered and
paramagnetic states remain an open question.

In addition to the bond order, there is a {\em nematic} long-range
order: every spin points along a common direction $\hat{\bf n}$ or its
opposite $-\hat{\bf n}$.  

Our large-$S$ analysis is in reasonable agreement with numerical
results for the $S=1/2$ planar pyrochlore antiferromagnet obtained by
Fouet {\em et al.}\cite{Fouet01} and Berg {\em et al.}\cite{Berg02}
They find a bond-ordered ground state of {\em almost} the same kind as
we do.  The ordering of the two sublattices of tetrahedra is no longer
independent: opposite patterns of the bond order are chosen.  The
ground state at $S=1/2$ appears to be a doubly degenerate spin
singlet.  Numerical data suggest the presence of a spin gap, which
seems to rule out the presence of spin order of the N\'eel or nematic
type.  This brings up the question of stability of the nematic order
that we have found at large $S$.  A recent calculation of
Canals,\cite{Canals02} based on the Dyson--Maleev approximation,
indicates that collinear ground states are locally stable at large $S$
but could become unstable below some critical value $S_c$.  Although
the Dyson--Maleev scheme is not a controlled approximation for small
$S$, Canals' scenario is consistent with the results reported here and
in Refs.~\onlinecite{Fouet01,Sindzingre02,Berg02}.

A lack of symmetry between the first and second-neighbor bonds
compells one to study a more general system with unequal first and
second-neighbor exchanges $J_1 \neq J_2$.  For large $S$, the
deviation of $j = J_2/J_1$ from the critical value $j_c = 1$ plays the
role of a ``magnetic field'' in the 3-state Potts model.  The 3
flavors correspond to the 3 collinear ground states of spins on a
tetrahedron (modulo a global rotation of the spins).  They can be
labelled by the location of frustrated bonds: diagonal, vertical, or
horizontal.  The ``magnetic field'' $h \propto J_2/J_1-1$ prefers the
``diagonal'' state when $J_2 < J_1$, in which case the ground state is
unique and the lattice symmetries are intact.  When $J_2>J_1$, the
``vertical'' and ``horizontal'' states are favored, which leads to a
spontaneous breaking of the rotational symmetry of the lattice.
Still, even at the critical point $J_2/J_1 = 1$ the symmetry between
the three flavors is not restored: the diagonal bonds remain different
from the horizontal and vertical ones, so that the symmetry is still
${\bm Z}_2$, rather than ${\bm S}_3$.  Accordingly, the order
parameter is of the Ising, rather than the 3-state Potts, model.  Put
another way, only one component of the Potts order parameter is used:
that which is orthogonal to the direction of the ``magnetic field''
$h$.  

At large $S$, N\'eel order is present on both sides of the critical
point $J_2/J_1 = 1$, but not at the critical point itself.  For $J_1 <
J_2$ (diagonal chains with a weak frustrated coupling) the N\'eel
order is induced by quantum fluctuations of spins.  There are signs
that, for a sufficiently weak interchain coupling, $J_{1} < J_{1c} =
{\cal O}(J_2 \sqrt{S} e^{-S/2})$, the N\'eel order may be destroyed.
The fate of the bond order is unknown, although chances are that it is
less susceptible to long-wavelength spin fluctuations.  

While it may be unreasonable to expect quantitative information about
small spin values from a $1/S$ expansion, we were tempted to make some
general statements about the observed behavior of the $S=1/2$
system.\cite{Fouet01,Sindzingre02,Berg02} It appears that there is some
family resemblance and that the zero-temperature phase diagrams for
large $S$ [Fig.~\ref{fig-phases}(a)] and for $S=1/2$
[Fig.~\ref{fig-phases}(b)] can be understood in similar terms, as far
as bond order is concerned.  We find it plausible that the bond
operators (rerpesented for $S=1/2$ by $2 \!\times\! 2$ Pauli matrices)
behave as spins of an Ising antiferromagnet in a transverse magnetic
field.  The Ising order parameter is an expectation value of $\tau_z$;
the ``magnetic field'' $h$ couples to $\tau_x$.  In zero transverse
field, there is a bond order of the antiferro type, as observed by
Fouet {\em et al.}  A nonzero transverse field $h \propto J_2/J_1-j_c$
induces quantum fluctuations of the ``spins'' $\tau_z$.  At some
critical value of the ``field'' the bond order melts and the ordered
phase has a finite extent, in agreement with the numerical work of
Sindzingre {\em et al.}\cite{Sindzingre02}

It is rather intriguing to find a valence-bond solid without N\'eel
order in the limit of large $S$.  This result is probably not unique
to the planar pyrochlore and we intend to pursue this avenue further
by studying other lattices.  In particular, it would be interesting to
solve a similar problem on the three-dimensional pyrochlore lattice
and determine the resulting phase.  In that case, the $S_3$ symmetry
of the Potts states is intact and the outcome should be different.
There are further open questions.  Are there lattices that contain
valence-bond {\em liquids} at large $S$?  What happens at intermediate
values of $S$?  We look forward to finding out.

\begin{acknowledgments}

We thank C. L. Henley for sharing his unpublished results.  We are
grateful to C. L. Broholm, A.V. Chubukov, S. Sachdev and S. L. Sondhi
for useful discussions, and to P. Sindzingre for a careful reading of
the manuscript.  We acknowledge hospitality of the Lorentz Centre of
Leiden University where part of this work was done.  Financial support
was provided in part by the Alfred P. Sloan Foundation (A.G.A.),
Research Corporation Award No. CC5491 (O.A.S.), and the NSF Grant
No. DMR-9978074 (O.T.).

\end{acknowledgments}

\appendix*

\section{Order from disorder in the chain limit}
\label{sec:app}

Here we derive the results presented in Sec.~\ref{sec-1st-order}.

\subsection{Notation}
 \label{sec:swave}

The starting point $J_1 = 0$ corresponds to completely decoupled
chains running along the diagonal directions $\hat{\xi}$ and
$\hat{\eta}$.  For technincal reasons, we will consider chains 
with an Ising anisotropy:
\begin{equation}
H_0 = J_2 \sum_i
\delta [S^x_i S^x_{i+1} + S^y_i S^y_{i+1}] + S^z_i S^z_{i+1}
\label{h_0}
\end{equation} 
Introduction of the anisotropy ($0 \leq \delta < 1$) helps stabilize
the collinear ground state of a single chain: at the Heisenberg point
($\delta = 1$) N\'eel order along the chain is destroyed by quantum
fluctuations.  At the technical level, this is caused by a divergent
$1/S$ correction to the staggered moment for $\delta =
1$.\cite{affleck} Then, strictly speaking, the initial assumption of
the long-range N\'eel order on the chain breaks down.  Therefore we
compute the effects of the interchain coupling at $\delta < 1$ and
then take the Heisenberg limit $\delta \to 1$.  Unlike the staggered
magnetization, the energy of zero-point motion (\ref{E-1}) does not
have an infrared divergence.  This justifies approaching the
Heisenberg limit $\delta \to 1$ from below.\cite{affleck} (Note,
however, that the interchain coupling is always taken to be of the
isotropic, Heisenberg kind.)

At the Ising point, $\delta=0$, the system has an extensive
degeneracy: there are $2^L$ ground states.  We parametrize them by
introducing a single Ising variable $\pm 1$ for every chain.  Then,
for example, on the $m$th chain along $\hat{\xi}$---whose spins have
coordinates ${\bf r} = (n+1/2,m)$---we have $S_{\bf r}^z=(-1)^{n+1}
s_m\, S$.  We will use the Holstein-Primakoff transformation keeping
the terms of orders $S, S^{1/2}$ and 1.  For a spin with $S^z > 0$,
\begin{eqnarray} 
S^z &=& S-a^\dagger a, \nonumber \\ S^+
&=&\sqrt{2S}\sqrt{1-\frac{a^\dagger a}{2S}}a = \sqrt{2S}a
+O\left(\frac{1}{\sqrt{S}}\right), \nonumber \\ S^-
&=&\sqrt{2S}a^\dagger \sqrt{1-\frac{a^\dagger a}{2S}} =
\sqrt{2S}a^\dagger +O\left(\frac{1}{\sqrt{S}}\right).  \nonumber
\end{eqnarray} 
For spins with $S_z < 0$ we rotate the reference frame about the
direction $\hat{\bf x}$ through $\pi$.  Then for a spin located on 
the $m$-th chain running along $\hat{\xi}$ we obtain
\begin{eqnarray}
\label{horiz}
\left(\begin{array}{c} S^x_{\bf r}\\ S^y_{\bf r}\\ S^z_{\bf r}
\end{array}
\right) &=& \left(\begin{array}{ccc }
      1 &0 &0  \\
      0 &(-1)^{n+1}s_m &0\\
      0 &0 &(-1)^{n+1}s_m
\end{array}
\right) \nonumber\\
&&\times
\left(\begin{array}{c }
      \frac{\sqrt{2S}}{2}(a_{\bf r} + a^\dagger_{\bf r})\\
      \frac{\sqrt{2S}}{2i}(a_{\bf r} - a^\dagger_{\bf r})\\
      S - a^\dagger_{\bf r} a_{\bf r}
\end{array}\right),
\end{eqnarray}
where ${\bf r} = (n+1/2,m)$. 

The effective potential generated by quantum fluctuations is a
function of $L$ N\'eel vectors $\hat{\bf n}_m$.  Invariance under
global spin rotations implies certain restrictions on the form of that
function.  For instance, a pairwise potential coupling chains with
staggered magnetizations $\hat{\bf n}_1$ and $\hat{\bf n}_2$ must be a
function of the scalar product $\hat{\bf n}_1 \cdot \hat{\bf n}_2$.
In addition, symmetry of the lattice requires---for chains running in
perpendicular directions--- that the coupling be invariant under
$\hat{\bf n}_1 \mapsto -\hat{\bf n}_1$, so that it must depend on
$(\hat{\bf n}_1 \cdot \hat{\bf n}_2)^2$.  With this in mind, we will
compute the lowest-order effect---${\cal O}(J_1^2)$---as a function of
$L(L-1)/2$ angles between staggered magnetizations $\theta_{mn} =
\arccos{(\hat{\bf n}_m \cdot \hat{\bf n}_n)}$.  The pairwise nature of
this potential allows us to tilt the spins uniformly on all $\eta$
chains.  It is convenient to choose $\hat{\bf n}_1$ as the $z$
direction and let $\hat{\bf n}_2$ lie in the $yz$ plane.

For potentials coupling more than two chains, one must consider a
general orientation of the staggered magnetizations involved.
However, because the leading term (\ref{2order}) already selects
collinear states, in higher orders we will work with collinear
configurations only.

Thus for the $n$th $\eta$ chain---whose spins reside at ${\bf r} =
(n,m+1/2)$---we perform an additional uniform rotation in the $yz$ plane:
\begin{eqnarray}
\label{vert}
\left(\begin{array}{c}
       S^x_{\bf r}\\
       S^y_{\bf r}\\
       S^z_{\bf r}
\end{array}
\right) &=& \left(\begin{array}{ccc }
      1 &0 &0  \\
      0 &(-1)^{m+1}t_n \cos\theta &(-1)^{m+1} t_n \sin\theta\\
      0 &-(-1)^{m+1}t_n \sin\theta &(-1)^{m+1}t_n \cos\theta
\end{array}
\right) \times \nonumber\\
&&\times \left(\begin{array}{c }
      \frac{\sqrt{2S}}{2}(a_{\bf r} + a^\dagger_{\bf r})\\
      \frac{\sqrt{2S}}{2i}(a_{\bf r} - a^\dagger_{\bf r})\\
      S - a^\dagger_{\bf r} a_{\bf r}
\end{array}\right)
\end{eqnarray}
Naturally, the intra-chain Hamiltonian is not affected by these
unitary rotations:
\begin{equation}
H_0 = J_2 S \sum_{\bf r} \left[ 2a^\dagger_{\bf r} a_{\bf r} 
+ \delta ( a_{\bf r} a_{\bf r + \hat{\xi}} 
+ a^\dagger_{\bf r} a^\dagger_{\bf r + \hat{\xi}}) \right]
\label{harm}
\end{equation}
for the $\xi$ chain.  After a Fourier transform,
\begin{equation}
H_0 = J_2 S\sum_p \left[ 2a^\dagger_p a_p + 
\delta \cos{p} (a_p a_{-p} + a^\dagger_p a^\dagger_{-p})\right],
\label{harmk}
\end{equation}
where $p$ is the lattice momentum along the chain direction.
Thermal Green's functions are given by
\begin{eqnarray}
    \langle a^\dagger(\omega, p)\, a(-\omega, -p)\rangle &=&
    \frac{\omega_0-i\omega}{\omega^2+\epsilon_p^2}
 \\
    \langle a(\omega, p)\, a(-\omega,-p)\rangle &=& 
    -\frac{\omega_0\gamma_p}{\omega^2+\epsilon_p^2},
\end{eqnarray}
where $\omega$ is a bosonic Matsubara frequency and
\begin{eqnarray}
    \omega_0 &=& 2J_2S,
 \nonumber \\
    \gamma_p &=&\delta \cos{p},
  \\
    \epsilon_p &=& \omega_0\sqrt{1 - \gamma_p^2}.
 \nonumber    
\end{eqnarray} 

The perturbation term in the Hamiltonian (for which we introduce no 
anisotropy)
\begin{equation}
V = \sum_{n,m} J_1 ({\bf S}_{n-1/2,m} + {\bf S}_{n+1/2,m}) 
\cdot
({\bf S}_{n,m-1/2} + {\bf S}_{n,m+1/2})
\label{interchain}
\end{equation}
couples linear combinations of spins ${\bf S}_{n-1/2,m} + {\bf
S}_{n+1/2,m}$ and ${\bf S}_{n,m-1/2} + {\bf S}_{n,m+1/2}$.  Note that
both linear combinations live on the same tetrahedron centered at
$(n,m)$.  It is convenient to introduce variables $\xi_{nm}$ and
$\eta_{nm}$ representing transverse spin fluctuations on the
respective diagonal links of the tetrahedron.  For a link along
$\hat{\xi}$,
\begin{equation}
    {\bf S}_{n-\frac{1}{2},m} + {\bf S}_{n+\frac{1}{2},m}
    = 2\sqrt{S}\left(\begin{array}{c }
      \xi^x_{nm}  \\
     s_m (-1)^{n+1} \xi^y_{nm}\\
      s_m (-1)^{n+1} \xi^z_{nm}
\end{array}\right).
\label{hv}
\end{equation}
By direct comparison to (\ref{horiz}) we obtain
\begin{eqnarray}
    \xi^x_{nm} &=& \frac{1}{2\sqrt{2}}
    (a_{n-\frac{1}{2},m}^\dagger+a_{n-\frac{1}{2},m} 
    +a_{n+\frac{1}{2},m}^\dagger +a_{n+\frac{1}{2},m}),
 \nonumber \\
    \xi^y_{nm} &=& \frac{i}{2\sqrt{2}}
    (a_{n-\frac{1}{2},m}^\dagger-a_{n-\frac{1}{2},m} 
    -a_{n+\frac{1}{2},m}^\dagger +a_{n+\frac{1}{2},m}),
 \nonumber \\
    \xi^z_{nm} &=& 0.
 \label{horizh} 
\end{eqnarray}
The longitudinal component $\xi^{z}$ is of order $S^{-1/2}$ and
has therefore been dropped at the current level of approximation.  
For $\eta$ chains we similarly define $\eta^a$ fields via
\begin{eqnarray}
    \eta^x_{nm} &=& \frac{1}{2\sqrt{2}}
    (a_{n,m-\frac{1}{2}}^\dagger+a_{n,m-\frac{1}{2}} 
    +a_{n,m+\frac{1}{2}}^\dagger +a_{n,m+\frac{1}{2}}),
 \nonumber \\
    \eta^y_{nm} &=& \frac{i}{2\sqrt{2}}
    (a_{n,m-\frac{1}{2}}^\dagger-a_{n,m-\frac{1}{2}} 
    -a_{n,m+\frac{1}{2}}^\dagger +a_{n,m+\frac{1}{2}}),
 \nonumber \\
    \eta^z_{nm} &=& 0.
 \label{vertv} 
\end{eqnarray}
The sum of the spins along an $\eta$ link is then expressed, according
to (\ref{vert}), as
\begin{eqnarray}
{\bf S}_{n,m-\frac{1}{2}} + {\bf S}_{n,m+\frac{1}{2}}=2\sqrt{S} 
\left(\begin{array}{ c}
\eta^x_{nm} \\
(-1)^{m+1} t_n \cos\theta \, \eta^y_{nm}  \\
(-1)^{m} t_n \sin\theta \, \eta^y_{nm} 
\end{array}\right),
 \label{vv}
\end{eqnarray}
The interchain coupling (\ref{interchain}) now reads
\begin{equation}
V = 4J_1S\sum_{n,m} (\xi^x_{nm} \eta^x_{nm} + 
\bar{\xi}^y_{nm} \bar{\eta}^y_{nm}),
\label{V-xi-eta}
\end{equation}
where we have introduced a shorthand notation
\begin{equation}
\bar{\xi}^y_{nm} = (-1)^n s_m \xi^y_{nm},
\hskip 5mm
\bar{\eta}^y_{nm} = (-1)^{m} t_n \cos\theta \ \eta^y_{nm}. 
\label{eq-bar}
\end{equation}

To complete the preparation stage we work out Green's functions of the
$\xi$ and $\eta$ variables for uncoupled chains:
\begin{eqnarray}
   \langle \xi_{nm}^a(\tau) \xi_{n'm'}^b(\tau')\rangle
    &=& \delta_{mm'} G^{ab}(\tau-\tau',n-n'),
 \\
   \langle \eta_{nm}^a(\tau) \eta_{n'm'}^b(\tau')\rangle
    &=& \delta_{nn'} G^{ab}(\tau-\tau',m-m'),
\end{eqnarray}
where indices $a$ and $b$ take on values $x$ and $y$.  The space-time
Green's functions $G^{ab}(\tau,n)$ are easily obtained in terms of their
Fourier transforms,
\begin{equation}
G^{ab}(\tau,n) = \frac{1}{\beta}\sum_{\omega} 
\int_{-\pi}^{\pi} \frac{dp}{2\pi} \, 
G^{ab}(\omega, p) \, e^{-i\omega\tau + ipn}
\end{equation}
which are given by the matrix
\begin{equation}
    \hat{G}(\omega, p) = 
\left( \begin{array}{cc}
     \frac{\omega_0(1-\gamma_p)\, \cos^2{(p/2)}}{\omega^2+\epsilon_p^2}
    & \frac{-i\omega\, \sin{(p/2)}\cos{(p/2)}}{\omega^2+\epsilon_p^2}
    \\ \frac{-i\omega\, \sin{(p/2)}\cos{(p/2)}}{\omega^2+\epsilon_p^2}
    & \frac{\omega_0(1+\gamma_p)\, \sin^2{(p/2)}}{\omega^2+\epsilon_p^2}
    \end{array} \right).
\label{G-p}
\end{equation}
Partial Fourier transforms $G^{ab}(\omega,n)$ are real and satisfy the
following identities:
\begin{eqnarray}
    &&G^{xx}(\omega,n) = (-1)^n G^{yy}(\omega,n) = G^{xx}(\omega,-n),
 \nonumber \\
    &&G^{xy}(\omega,n) = G^{yx}(\omega,n) 
    =  -G^{xy}(\omega,-n), 
\nonumber \\
&&G^{xy}(\omega,n) = (-1)^{n+1} G^{xy}(\omega,n).
 \label{relx}
\end{eqnarray}
The last line suggests that the off-diagonal components vanish for
{\em even} distances $n$.  At the Heisenberg point, the diagonal
components vanish for {\em odd} distances.  

\subsection{Order $J_1^2$}

The first nonvanishing correction to the free energy comes at the
second order in $J_1^2$ and can be expressed as the second moment of
the Eucledian action:
\begin{eqnarray}
E^{(2)} = -\frac{1}{2! \beta} 
\left\langle \left( \int_0^\beta d\tau \ V \right)^2\right\rangle.
\end{eqnarray}
The perturbation $V$ given by Eq.~(\ref{V-xi-eta}).  Its second moment 
contains quartic averages 
\begin{equation}
\left\langle 
(\xi^x_{nm} \eta^x_{nm} + \bar\xi^y_{nm} \bar\eta^y_{nm})
(\xi^x_{n'm'} \eta^x_{n'm'} + \bar\xi^y_{n'm'} \bar\eta^y_{n'm'})
\right\rangle.
\end{equation}
(Time variables are omitted for brevity.)  Dependence on the
staggered magnetizations $s_m$ and $t_n$ and the tilting angle
$\theta$ comes through the $y$ components---see Eq.~(\ref{eq-bar}).
Furthermore, terms containing two $x$ and two $y$ components vanish by
symmetry: they have one factor of $s$ only and staggered
magnetizations $s_m$ are odd under reflections in any line of spins
along $\eta$.  Therefore the only possibility is four $y$ components:
\begin{eqnarray}
\left\langle 
\bar\xi^y_{nm} \bar\eta^y_{nm}
\bar\xi^y_{n'm'}  \bar\eta^y_{n'm'} 
\right\rangle = (-1)^{m+m'+n+n'}\cos^2{\theta}
\nonumber\\
\times 
s_m s_{m'} t_{n} t_{n'} 
\left\langle \xi^y_{nm} \xi^y_{n'm'} \right\rangle
\left\langle \eta^y_{nm} \eta^y_{n'm'} \right\rangle,
\end{eqnarray}
where we have used Gaussian statistics of the variables $\xi$ and
$\eta$.  It is evident that lattice points $(nm)$ and $(n'm')$ must
belong to a $\xi$ chain and an $\eta$ chain simultaneously, so that
they are the same point.  This term therefore gives rise to a contact
interaction of crossing chains.  The correction to the energy of a
ground state is found by taking the limit of zero temperature ($\beta
\to \infty$):
\begin{eqnarray}
E^{(2)} &=& -\frac{(4 J_1 S)^2}{2!} 
\int \frac{d\omega}{2\pi} \ 
[G^{yy}(\omega,0)]^2 \sum_{mn} \cos^2{\theta}
\nonumber\\
&=& -I(\delta) \frac{J_1^2 S}{J_2} 
\sum_{mn} \left(\hat{\bf n}_m \cdot \hat{\bf n}_n\right)^2,
\label{eq-2-order}
\end{eqnarray}
where $\hat{\bf n}_m$ and $\hat{\bf n}_n$ are the directions of
staggered magnetizations on chains running in the $\xi$ and $\eta$
directions, respectively; $G(\omega,0)$ is the real-space Green's
function $G(\omega,n)$ at distance $n=0$.  The numerical constant
$I(\delta)$ is given here for the Ising ($\delta=0$) and Heisenberg
($\delta = 1)$ limits:
\begin{equation}
I(0) = \frac{1}{4}, 
\hskip 5mm
I(1) = \frac{4-\pi}{2\pi}.
\end{equation}

Eq.~(\ref{eq-2-order}) constitutes an order-by-disorder effect:
collinear spin configurations ($\hat{\bf n}_m = \pm \hat{\bf n}_n$)
minimize the energy of quantum fluctuations.  There are $2^L$ such
collinear states in an $L\times L$ lattice with periodic boundary
conditions.  Their degeneracy is partially lifted at the order
$J_1^4$, as we discuss next.
 
\subsection{Order $J_1^4$}

\begin{figure}
\includegraphics[width=0.9\columnwidth]{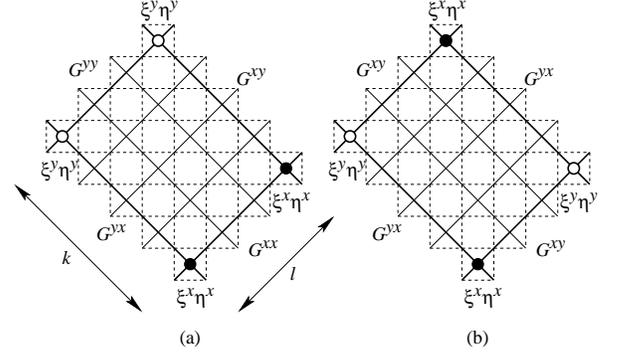} 
\caption{ Computation of the quantum correction to the free energy at
${\cal O}(J_1^4)$.  Filled dots reprsent vertices $\xi^x_{mn}
\eta^x_{mn}$, open ones $\xi^y_{mn} \eta^y_{mn}$; the latter
contribute a factor $(-1)^{m+n} s_m t_n$ to the diagram.  (a) The
first term in Eq.~(\ref{j4}).  The diagram has a prefactor $(-1)^l t_n
t_{n+l}$.  (b) The third term in Eq.~(\ref{j4}).  Prefactor 
$(-1)^{k+l} s_m s_{m+k}\, t_n t_{n+l}$.  }
\label{fig-j4}
\end{figure}

In the remainder of this section we will consider collinear vacua.
Therefore the unitary transformation matrix in Eq.~(\ref{vert})
becomes diagonal and (\ref{vv}) acquires the same form as (\ref{hv}).
The ground-state energy will depend on the Ising variables $s$ and
$t$.  The first non-trivial correction to the energy comes from the
term of the fourth order in $J_1$:
\begin{equation}
E^{(4)} = -\frac{1}{4! \beta}
\left\langle \left(\int d\tau\, V\right)^4 \right\rangle.
\end{equation}
The dependence on staggered magnetizations $s$ and $t$ comes through
the vertex $\bar\xi^y \bar\eta^y$ (\ref{eq-bar}).  Symmetry arguments
used in the preceding calculation show that $s$ and $t$-dependent
diagrams must contain exactly two vertices $\xi^x \eta^x$ and two
$\bar\xi^y \bar\eta^y$.  We obtain
\begin{widetext}
\begin{eqnarray}
E^{(4)} =
- \, 2 \times 6 \times \frac{(4 J_1 S)^4}{4! \beta} \int   
\frac{d\omega}{2\pi}
\sum_{k,l,m,n}
\big[
(-1)^{k} s_{m} s_{m+k} \,
G^{xy}(\omega,l) G^{yy}(\omega,k) G^{yx}(\omega,-l) G^{xx}(\omega,-k)
\nonumber \\
+ (-1)^l t_{n}t_{n+l} \,
G^{xx}(\omega,l) G^{xy}(\omega,k) G^{yy}(\omega,-l) G^{yx}(\omega,-k)
\nonumber \\ 
 + (-1)^{l+k} s_{m}s_{m+k} \, t_{n}t_{n+l} \,
G^{xy}(\omega,l) G^{yx}(\omega,k) G^{xy}(\omega,-l) G^{yx}(\omega,-k)
\big].
\label{j4}
\end{eqnarray}
\end{widetext}
Feynman diagrams contributing to Eq.(\ref{j4}) consist of a
rectangular path formed by two $\xi$ chains and two $\eta$ chains
(Fig.~\ref{fig-j4}).  The factors $2 \times 6$ have a
combinatorial origin.  At this order, the interchain coupling $J_1$ 
generates a two-chain interaction (between parallel chains) and a 
four-chain one (involving four crossing chains):
\begin{eqnarray}
E^{(4)} = \frac{2J_1^4 S}{ J_2^3} 
\sum_{l>0} \sum_{k>0}
\sum_{m} \sum_{n}
(A_{k} \, s_{m} s_{m+k} 
\label{eq-A-B} 
\\
+ A_{l} \, t_{n} t_{n+l}
- B_{kl} \, s_{m} s_{m+k} t_{n} t_{n+l}),
\nonumber
\end{eqnarray}
Dimensionless couplings $A_l \geq 0$ and $B_{kl} \geq 0$ are
\begin{eqnarray}
    A_k &=& 2^7 (J_2S)^3 \int\frac{d\omega}{2\pi} 
[G^{xx}(\omega,k)]^2 \sum_{l}[G^{xy}(\omega, l)]^2,
 \nonumber \\
    B_{kl} &=& 2^8 (J_2S)^3 \int\frac{d\omega}{2\pi}
    [G^{xy}(\omega,k)]^2 [G^{xy}(\omega,l)]^2.
 \label{ABcoeff}
\end{eqnarray}
They fall off quickly with the distances $k$ and $l$.  We have used
the properties of the Green's functions (\ref{relx}) in deriving them.
Below we discuss the Ising ($\delta = 0$) and Heisenberg ($\delta
\to 1$) limits.

\subsubsection{Ising limit}
\label{sec:swave-ising}

In the Ising limit, the magnons $\xi$ and $\eta$ have an infinite mass
and are unable to move far along the chain.  Therefore $G^{ab}(\omega,
l) = 0$ for $|l|>1$.  As a result, the effective potential depends on
the nearest-neighbor products $s_m s_{m+1}$ and $t_m t_{m+1}$:
\begin{eqnarray}
E^{(4)}_{\rm Ising}
= \frac{1}{128} \frac{J_1^4 S}{J_2^3} \sum_{m}\sum_{n} 
\left(s_{m}s_{m+1} + t_{n}t_{n+1}\right.  
\nonumber\\
- \left. s_{m}s_{m+1} t_{n}t_{n+1}  \right). 
\label{Ien} 
\end{eqnarray}
It is minimized by the ground states (\ref{chain-af}), in which
staggered magnetizations have opposite signs on neighboring parallel
chains.  Such configurations minimize every term in Eq.~(\ref{Ien}).

\subsubsection{Heisenberg limit}
\label{sec:swave-heis}

At the Heisenberg point, diagonal (off-diagonal) components of the
propagator (\ref{G-p}) contain even (odd) harmonics of the translation
operator $e^{ik}$ only.  In real-space terms, a magnon can travel an
even distance by keeping its polarization, an odd distance by flipping
it.  Thus pairwise chain interactions are induced for parallel chains 
with an even separation; quartic ones involve pairs of parallel chains 
with an odd separation:
\begin{eqnarray}
E^{(4)} &=&
\frac{2J_1^4 S}{ J_2^3} 
\sum_{k=1}^{\infty} \sum_{l=1}^{\infty} \sum_{m} \sum_{n} 
(A_{2k} s_{m}s_{m+2k} 
\label{Isotren}\\
&+& A_{2l} t_{m}t_{m+2l}
-B_{2k-1,2l-1}\, s_{m}s_{m+2k-1}\, t_{n}t_{n+2l-1} ).
\nonumber
\end{eqnarray}
The dimensionless coefficients $A$ and $B$ are given below.  As
previously noted, this potential is invariant under a reversal of
spins on every other diagonal chain (\ref{staggering}).  We will
discuss the origin of this dynamical symmetry in Sec.~\ref{sec-chains}
and demonstrate that it holds to an arbitrary order in $J_1$.

The dominant term in (\ref{Isotren})---proportional to
$B_{1,1}$---describes a coupling of four chains intersecting around an
open square.  It is minimized by spin configurations in which diagonal
chains running in the same direction have either constant staggered
magnetizations (\ref{chain-f}), or alternating ones (\ref{chain-af}).
These states---shown in Fig.~\ref{fig-cboard-vacua} (b) through
(e)---are related by a staggering transformation (\ref{staggering}).
Therefore they remain degenerate even upon the inclusion of all two-
and four-chain interactions in Eq.~(\ref{Isotren}).

The second-largest term---a two-chain potential proportional to
$A_2$---is minimized by N\'eel states of a different kind: the ones
with opposite staggered magnetizations for {\em second}-neighbor
chains.  Thus we obtain another viable candidate for a ground state:
\begin{equation}
s_{4n} = s_{4n+1} = -s_{4n+2} = -s_{4n+3}
\label{stripe}
\end{equation}
and similarly for $t$.  

Because the fourth-order energy correction (\ref{4order}) contains
oscillating terms, it is best to evaluate the energies at order
$J_1^4$ exactly.  We find that the energy of the states
(\ref{chain-f}) and (\ref{chain-af}) is indeed lower by $C J_1^4
S/J_2^3$ per spin, where
\begin{equation}
C = \frac{\sqrt{2}}{4} - \frac{45}{128} \approx 1.991 \times 10^{-3}.
\label{eq-C}
\end{equation}

\subsection{Coupling coefficients $A_l$ and $B_{kl}$}

In what follows we specialize to the Heisenberg case, $\delta = 1$.
Coefficients (\ref{ABcoeff}) of the two- and four-chains interaction
(\ref{4order}) are given in terms of the following integrals
\begin{eqnarray}
G^{xx}(\omega,n) &=& \frac{\omega_0}{2} \int_{-\pi}^\pi \frac{dp}{2\pi}\, 
   \frac{\sin^2{p} \, e^{ipn}}{\omega^2+\omega_0^2 \sin^2{p}},
 \nonumber \\
   G^{xy}(\omega,n) &=& \frac{\omega}{2i} \int_{-\pi}^\pi \frac{dp}{2\pi}\, 
   \frac{\sin p \, e^{ipn}}{\omega^2 + \omega_0^2 \sin^2{p}}.
 \label{integralsdef}
\end{eqnarray}

Introduce an auxiliary variable $u$:
\begin{equation}
\omega = \omega_0 \sinh{u}.  
\end{equation}
We find:
\begin{eqnarray}
G^{xx}(\omega,n) = &
\left\{
\begin{array}{ll}
\displaystyle{\frac{\delta_{n0} - \tanh{|u|}}{2\omega_0}\, e^{-|nu|} } 
& n \mbox{ is even,} \\
0 & n \mbox{ is odd,}
\end{array}
\right.
\end{eqnarray}
where $\delta_{mn}$ is the Kronecker delta;
\begin{eqnarray}
G^{xy}(\omega,n) = &
\left\{
\begin{array}{ll}
0 & n \mbox{ is even,} \\
\displaystyle{\frac{\tanh{u}}{2\omega_0}\, e^{-|nu|}\, {\rm sgn}\, n}
& n \mbox{ is odd.}
\end{array}
\right.
\end{eqnarray}
\begin{equation}
\sum_{n} \left[G^{xy}(\omega,n)\right]^2 
= \frac{\tanh{|u|}}{8\omega_0^2 \cosh^2{u}}.
\end{equation}
Substituting these into Eq.~(\ref{ABcoeff}) produces the following
expressions for nonvanishing coupling constants in the effective
potential (\ref{eq-A-B}):
\begin{eqnarray}
& \displaystyle{
A_{l} = 
\int_0^1 \frac{dx}{2\pi}\, \frac{(1-x)^3\, x^{|l|-1/2}}{(1+x)^4},
}
& \mbox{$l$ is even,} 
\label{e/o}
\\
& \displaystyle{
B_{kl} = 
\int_0^1 \frac{dx}{2\pi}\, \frac{(1-x)^4\, x^{|k|+|l|-3/2}}{(1+x)^3},
} 
& \mbox{$k$ and $l$ are odd.} 
\nonumber
\end{eqnarray}
Numerical values for the first few terms are:
\begin{eqnarray} 
&A_{2} = 1.41 \times 10^{-3}, 
&B_{1,1} = 6.86 \times 10^{-3},
\nonumber\\
&A_{4} = 1.63 \times 10^{-4}, 
&B_{1,3} = 3.47 \times 10^{-4},
\\
&A_{6} = 3.84 \times 10^{-5}, 
&B_{1,5} = B_{3,3} = 5.29 \times 10^{-5}.
\nonumber
\end{eqnarray}

Although these coupling constants fall off rather quickly---as
$|l|^{-4}$ and $(|k|+|l|)^{-5}$ at large distances---partial
cancellations in Eq.~(\ref{eq-A-B}) call for a careful comparison of
energies of the candidate ground states.  In a state with equal
staggered magnetizations on all chains (\ref{chain-f}), all products
$s_m s_{m+k} = t_m t_{m+l} = 1$.  The fourth-order correction to the
energy is
\begin{equation}
\frac{E_{\rm ++++}^{(4)}}{N_{\rm spins}} 
= \frac{J_1^4 S}{J_2^3} 
\left(
2\sum_{l=1}^{\infty} A_{2l} 
- \sum_{k=1}^{\infty} \sum_{l=1}^{\infty} B_{2k-1,2l-1}
\right)
\end{equation}
per spin.  The state with alternating staggered magnetizations
(\ref{chain-af}) is degenerate with it by virtue of the staggering
symmetry (\ref{staggering}).  In the $(++--)$ state (\ref{stripe}), $s_m
s_{m+k}$ oscillates as a function of $m$ (and thus averages out to
zero) for an odd $k$ and equals $(-1)^{k/2}$ for an even $k$.  Hence
the expression for the correction to the ground-state energy at this
order:
\begin{equation}
\frac{E_{++--}^{(4)}}{N_{\rm spins}} 
= \frac{J_1^4 S}{J_2^3} 
\times 2\sum_{l=1}^{\infty} A_{2l} (-1)^l.
\end{equation}
The energy difference is then 
\begin{equation}
\frac{E_{++--}^{(4)} - E_{++++}^{(4)}}{N_{\rm spins}} 
= C \, \frac{J_1^4 S}{J_2^3} > 0,
\end{equation}
where the constant is 
\begin{eqnarray}
C &=& \int_0^1 \frac{dx}{2\pi} \, 
\frac{(1-x)^2 \, (1-4x+x^2) \, x^{1/2}}{(1+x)^5 \, (1+x^2)}
\nonumber\\
&=& \frac{\sqrt{2}}{4} - \frac{45}{128} \approx 1.991 \times 10^{-3}.
\end{eqnarray}

\end{document}